\def\kms{\ifmmode{\,\hbox{km}\,s^{-1}}\else {\rm\,km\,s$^{-1}$}\fi}

\def\msun{{\rm\,M_\odot}}
\def\lsun{{\rm\,L_\odot}}

\def\kmsm{{\rm\,km\,s^{-1}\,Mpc^{-1}}}
\def\kmps{{\rm\,km\,s^{-1}}}
\def\hmpc{\ifmmode{h^{-1}\,\hbox{Mpc}}\else{$h^{-1}$\thinspace Mpc}\fi}
\def\hkpc{$h^{-1}$\thinspace kpc}
\def\eg{{\it e.g.}~}

\def\et{{\it et~al.}~}

\def\ie{{\it i.e.}~}
\def\sigp{\ifmmode{\sigma_p}\else {$\sigma_p$}\fi}
\def\sig1{\ifmmode{\sigma_1}\else {$\sigma_1$}\fi}
\def\r200{\ifmmode{r_{200}}\else {$r_{200}$}\fi}
\def\rp200{\ifmmode{r^\prime_{200}}\else {$r^\prime_{200}$}\fi}

\def\mlobs{289\pm50}
\def\mlcor{213\pm59}
\def\mlclose{1136\pm138}
\def\omlz{0.25\pm0.05} 

\def\omzeroc{0.19\pm0.06}
\def\systerr{\pm0.04}
\def\afit{0.66\pm0.09}
\def\mvadj{0.82\pm0.14}
\def\ladj{0.11\pm0.05} 
\def\sigall{1.05\pm0.04}

\documentstyle[11pt,aaspp4]{article}
\tighten

\slugcomment{DRAFT: \today}

\begin{document}

\title
{The Average Mass and Light Profiles of Galaxy Clusters}

\author
{
R.~G.~Carlberg\altaffilmark{1,2},
H.~K.~C.~Yee\altaffilmark{1,2},
\& E.~Ellingson\altaffilmark{1,3}
}

\altaffiltext{1}{Visiting Astronomer, Canada--France--Hawaii Telescope, 
        which is operated by the National Research Council of Canada,
        le Centre National de Recherche Scientifique, and the University of
        Hawaii.}
\altaffiltext{2}{Department of Astronomy, University of Toronto, 
	Toronto ON M5S~3H8 Canada}
\altaffiltext{3}{Center for Astrophysics \& Space Astronomy,
	University of Colorado, CO 80309, USA}


\begin{abstract}
The systematic errors in the virial mass-to-light ratio, $M_v/L$, of
galaxy clusters as an estimator of the field $M/L$ value are assessed.
We overlay 14 clusters in redshift space to create an ensemble cluster
which averages over substructure and asymmetries.  The combined
sample, including background, contains about 1150 galaxies, extending
to a projected radius of about twice \r200.  The radius \r200, defined as
where the mean interior density is 200 times the critical density, is
expected to contain the bulk of the virialized cluster mass.  The
dynamically derived $M(\r200)/L(\r200)$ of the ensemble is $(\mvadj)
\langle M_v/L \rangle$.  The $M_v/L$ overestimate is attributed to not
taking into account the surface pressure term in the virial equation.
Under the assumption that the velocity anisotropy parameter is in the
range $0\le\beta\le\twothirds$, the galaxy distribution accurately
traces the mass profile beyond about the central $0.3\r200$.  There
are no color or luminosity gradients in the galaxy population beyond
$2\r200$, but there is $\ladj$ mag fading in the $r$ band luminosities
between the field and cluster galaxies. We correct the cluster virial
mass-to-light ratio, $M_v/L=\mlobs h\msun/\lsun$ (calculated assuming
$q_0=0.1$), for the biases in $M_v$ and mean luminosity to estimate
the field $M/L=\mlcor h\msun/\lsun$.  With our self-consistently
derived field luminosity density, $j/\rho_c=\mlclose h\msun/\lsun$ (at
$z\simeq\onethird$), the corrected $M/L$ indicates
$\Omega_0=\omzeroc\systerr$ (formal $1\sigma$ random error and
estimated potential systematic errors) for those components of the
mass field in rich clusters.
\end{abstract}

\clearpage
\section{Introduction}

The existence of dark matter was discovered in galaxy clusters where
the velocity dispersions are nearly an order of magnitude higher than
expected from the gravitational binding provided by the stellar masses
of their visible galaxies (\cite{z33,smith,z37,schwarz}). Clusters are
gravitationally bound, quasi-equilibrium systems assembled over a
Hubble time via the infall of the mass in the surrounding field
(\cite{vanA,abellrev,peebles_coma,gg,white_coma,fg,bert}), which
implies that both the dark matter and the galaxies within clusters
have their origins in the field. Because clusters are large systems
that draw their mass and galaxy content from regions 20\hmpc\ across,
measurements of cluster $M/L$ values should be representative of the
field value, although not necessarily identical to it because of
differential galaxy evolution.  The product of the field $M/L$ with
the field luminosity density, $j$, is equal to the mean mass density
of the universe, $\rho_0$ (\cite{oort}).  The cosmological density
parameter, $\Omega_0\equiv\rho_0/\rho_c$, is therefore estimated as
the ratio of the cluster $M/L$ (corrected to the field) to the
$(M/L)_c\equiv\rho_c/j$ for closure (\eg\
\cite{gunn}).  The resulting $\Omega$ estimate has no dependence on
$H_0$ for dynamically measured cluster masses (\cite{gt}).  The
purpose of this paper is mainly to correct our cluster virial
mass-to-light ratio, $M_v/L$, (Carlberg, \et\ 1996, hereafter
\cite{global}) to the field $M/L$. 

Cluster dynamical masses are usually calculated from the virial mass
estimator, $M_v$, which has the drawback that it makes explicit assumptions
about the dynamical state of the clusters. Its reliability is
critically dependent on the galaxy population being in dynamical
equilibrium with the cluster potential and the galaxy distribution
having the same spatial distribution as the total mass distribution of
the cluster.  Measured virial mass-to-light ratios of clusters,
$M_v/L$, are generally in the range $200-600h\msun/\lsun$ (\eg\
\cite{gunn,rgh}) which, in ratio to $(M/L)_c\simeq1500h$
(\cite{eep,loveday}) indicates $\Omega\simeq0.1-0.4$. Cluster $\Omega_0$
values have not generally been accepted as conclusive because there
are a number of uncontrolled sources of possible systematic error.
For instance, cluster galaxies with blue colors are known to have a
higher velocity dispersion and are more extended than the red galaxies
(\cite{rpkk,kg}) leading to virial mass estimates that are about a factor of
four different for our dataset.  The substantial color difference
between cluster and field galaxies opens up the possibility that their
luminosities per unit mass are quite different.  The assumptions which
go into the cluster $\Omega$ calculation need to be tested for their
validity for a specific sample.

The possibility that the mass distributions of galaxy clusters are
more extended than their constituent galaxy populations has been
recognized for many years (\cite{limber}). The detection problem is that
dynamical mass estimators do not have any useful sensitivity to
cluster mass outside the orbits of the galaxies.  A specific situation
which leads to this bias is found in N-body simulations of cluster
buildup in an $\Omega=1$ cosmology with a realistic amount of
substructure (\cite{wr,cd,vbias,kw,vank}). In this case the simulated
``galaxies'' become more concentrated within the cluster than its
mass.  That is, the virial mass calculated from the ``galaxies''
systematically underestimates the total mass of the cluster in these
simulations by as much as order of magnitude.

There are some observational indications of the possibility of a
higher $\Omega$ than that calculated from virial masses.  Detailed mass
modeling for the Coma cluster gives a projected mass of
$5.8\times10^{15}h^{-1}\msun$ within $3^\circ$ of the center (with a
normal mass-to-light ratio, \cite{kg}), whereas the virial mass is
generally taken to be $2.1\times10^{15}h^{-1}\msun$ for the same
region.  However, this can also be ascribed to the substantial cluster
to cluster variations in the shape of the velocity dispersion profile
(\cite{hartog}) due to the asphericity and substructure within
individual clusters. Another indication comes from substructure, which
is a complication in the analysis of clusters as individuals, but can
be turned to advantage as an indirect, model dependent, $\Omega$
estimator. The results of substructure analysis remain controversial:
some workers favor $\Omega\simeq1$ (\cite{rlt,mohr}), while others
suggest much smaller $\Omega$ values (\cite{tsai_buote}).

The CNOC (Canadian Network for Observational Cosmology) cluster
redshift survey is designed to derive a cluster $\Omega_0$ within a
homogenous, self-contained, sample where we can explicitly test for
systematic errors. The catalogues contain $\sim$2600 redshifts in the
fields of 16 clusters (\cite{yec}) from which \cite{global}) derived
an average $\langle {M_v/L}\rangle=\mlobs h\msun/\lsun$ (calculated
with $q_0=0.1$). The clusters are consistent with having a universal
$M_v/L$ value (within the errors of about 25\%) independent of their
velocity dispersion, mean color of their galaxies, blue galaxy
content, or mean interior density.  The field galaxies in the dataset,
with the same corrections, over the same redshift range, yield the
closure mass-to-light ratio, $\rho_c/j$, to be $\mlclose
h\msun/\lsun$.  Consequently the virial mass-to-light ratio implies
$\Omega_0=\omlz$ at a mean redshift of 0.32, where the error is the
formal $1\sigma$ random error.  In \cite{global} the derived $\Omega$
was erroneously corrected to a redshift zero $\Omega_0$. The
calculation uses a co-moving volume element ($q_0$ dependent) and
Hubble constant which are both redshift zero values which do not need
further corrections.

Testing the accuracy of the virial mass was a primary consideration in
the design of the CNOC observations.  The data allow us to derive a
radially resolved projected number density profile, $\Sigma_N(R)$, and
a projected velocity dispersion, $\sigma_p(R)$ (in the following, $R$
is used for projected radius; whereas $r$ denotes the spherical
co-ordinate).  The sample meets the two main requirements for deriving
a radially resolved mass profile which are, first, accurate control of
the background, and second, enough clusters to average over the
aspherical complications of individual clusters.  In Section 2 we
review the properties of our sample. We define scaling velocities and
radii which we use in Section 3 to combine all but two ``binary''
clusters to create an ``ensemble cluster''. This averaging increases
the signal-to-noise ratio of our measurements, and is sufficient to
justify the assumption of an equilibrium, spherical, cluster in the
analysis.  The clusters are sampled to densities in redshift space
that are comparable to the field density, requiring a background
subtraction which is discussed in Section 4.  The mean surface density
profile of galaxies within the cluster is derived in Section 5, and
fitted with a spatial density profile. The projected velocity
dispersion profile is derived in Section 6, and fitted to a radial
velocity dispersion function for a variety of orbital shape
assumptions.  Using these results the mass-to-light profile is derived
from Jeans equation in Section~7.  In Section~8 we examine the
relative colors and luminosities of field and cluster galaxies as a
measurement of the differential evolution of the two populations. The
corrections and error budget are presented in Section 9, followed by
our conclusions and discussion in Section 10.

\section{Sample and Observations}

The CNOC sample is designed to create a dataset that allows as
complete internal control of all aspects of the cluster $\Omega$
estimate as possible. It is essential that the sample be useful for a
test of the equilibrium hypothesis and whether the virial mass is
biased in some way.  Furthermore the sample is used to measure
differential luminosity evolution between cluster and field galaxies.
On the basis of some n-body simulation data (\cite{cnoc1}) it was
argued that these could be best met within the constraints of
available observational resources with a set of a dozen or so clusters
at $z\simeq 1/3$ with a total of 3000 or so accurate redshifts.

The cluster sample was chosen from the Einstein Medium Sensitivity
Survey Catalogue of X-ray clusters (\cite{emss1,emss2,gl}) to have a
high X-ray luminosity, $L_x>4\times10^{44}$ erg~s$^{-1}$, which helps
guarantee that the clusters are reasonably virialized and have
relatively high masses, making them easier to study.  The clusters
chosen, Table~\ref{tab:r200}, are at moderate redshifts,
$0.17<z<0.55$, which has a number of advantages for mass
estimation. They are sufficiently distant that they have a significant
redshift interval over which the density of foreground and background
galaxies are nearly uniformly sampled in redshift.  At
$z\simeq\onethird$, a typical cluster diameter of $3$\hmpc\ (comoving)
spans an angle of about 12\arcmin, which is sufficiently small that
uniformity of photometry and sample selection is relatively easily
assured.  This angle is also comparable to the field size of the
Canada-France-Hawaii Telescope (CFHT) Multiple-Object-Spectrograph
(\cite{mos}), approximately 10\arcmin\ square.

Observations were made at CFHT in 24 assigned nights in 1993 and 1994,
of which 22 were usefully clear.  The sample and observational
techniques are described in detail elsewhere (\cite{yec}).  The
primary catalogues contain Gunn $r$ magnitudes and $g-r$ colors for
25,000 objects, of which about 2600 have velocities, on the average
accurate to about 100~\kms.  About one-third of the sample is cluster
galaxies. The other two-thirds are field galaxies, although less than
half of them are within the fair sample region defined by the
band-limiting filter's passband and a $\Delta z=0.01$ buffer zone
added to the upper and lower cluster redshift range.  For reasons of
observational efficiency, the survey is not ``complete'' in the usual
sense.  Hence, each object with a redshift has calculated weights
which are the inverse of the magnitude selection function and the
magnitude-dependent geometric selection function (\cite{yec}).  There
are small changes in the cluster dynamical parameters of
Table~\ref{tab:r200} from those in the global analysis
(\cite{global}). The values here are derived using the final survey
catalogues which accounts for the small changes from those given in
earlier papers.  All calculations in this paper assume $H_0=100\kmsm$,
$\Omega_0=0.2$ and $\Lambda=0$, although these choices do not affect
our main conclusions in any significant way.

\section{Dynamical Parameters of the Clusters}

The primary goal of this paper is to measure any systematic biases in
$M_v/L$ as an estimator of the field value. This is done by
independently measuring the mass and constraining differential
luminosity evolution of cluster and field galaxies. We adopt the
virial mass, 
\begin{equation}
M_v = {3\over G} \sigma_1^2 r_v,
\label{eq:virial}
\end{equation}
as the standard estimator, where \sig1\ and $r_v$ are defined below in
Equations~\ref{eq:sig} and \ref{eq:rv}.  The drawback of the virial
mass is that it is a statistically ``inefficient'' estimator
(\cite{bt81,bfg}). Its positive features are that it is completely
independent of the distribution of orbital shapes and it is not unduly
sensitive to background contamination. An alternate estimator is the
projected mass (\cite{bt81}), which we calculated but do not
report. Its main drawback is that it is quite sensitive to background
contamination of the data at large radii. For the standard velocity
ellipsoid shape the projected mass is always larger than the virial
mass, which we find below already overestimates the cluster mass.
Hence, we adopted the virial mass as our standard estimator.

The first step in testing the accuracy of $M_v$ is to construct a
low-noise, effectively spherical, ``ensemble'' cluster, which averages
over internal substructure and the variation of projected quantities
with viewing angle.  Our clusters span about a factor of two in
velocity dispersion and hence are not a uniform set of physically
identical objects.  Simply overlaying all the clusters onto a common
center of mass in physical velocity and position units has some
significant advantages (an approach that will be pursued in another
paper), but has the disadvantage that any gradients of density or
velocity dispersion tend to be blurred out.  The natural scaling
parameters are each cluster's RMS velocity dispersion and the
characteristic radius at which the cluster is expected to be in an
effective equilibrium, which turns out to be a function of the RMS
velocity dispersion, alone, within our approximation.

Deciding which galaxies in redshift space are cluster members is
fundamentally ambiguous.  That is, a cluster galaxy with a completely
reasonable line-of-sight velocity of $1000~\kmps$ appears in redshift
space at 10\hmpc\ from the cluster's center of mass, far outside
the virialized cluster and overlaying field galaxies.  This
complication increases in severity at large distances from the
cluster center, which our sample is specifically designed to
probe. One straightforward solution to this problem, which
we adopt, is to explicitly
subtract the mean density of field galaxies in redshift
space from the cluster distribution.

The internal kinetic energy of the cluster is calculated from the
characteristic velocity dispersion, $\sigma_1$. We iterate the
classical estimator
\begin{equation}
\sigma_1^2 = \left({\sum_i w_i}\right)^{-1} \sum_i w_i (\Delta v_i)^2, \qquad
\label{eq:sig}
\end{equation}
where the $\Delta v_i=c(z_i-\overline{z})/(1+\overline{z})$ are the
peculiar velocities in the frame of the cluster and $\overline{z}$ is
the weighted mean redshift of the cluster.  The weights are calculated
from the redshift and photometric catalogues to allow for statistical
incompleteness of sampling (\cite{yec}).  The key to the use of this
estimator is to have an objective choice of the galaxies that are
likely to be cluster members. The adopted method (\cite{global}) is as
follows. First a choice of the cluster redshift range is made from
which a trial \sig1\ is calculated. Then all galaxies between 5 and 15
\sig1\ are used to make a background estimate, which is subtracted
from the weights of the galaxies selected to be in the cluster. If the
\sig1\ calculated from the background subtracted weights is within the
68\% bootstrap confidence range of the trial \sig1\ (\cite{et}), then
the procedure is stopped; otherwise, the redshift limits are increased
or decreased to find a convergence.

Precisely the same galaxies as determined to be ``in the cluster''
from our \sig1\ calculation are given to the iterated bi-weight
estimator (\cite{bfg}). The resulting estimates of the velocity
dispersion are, on the average, $1.073\pm0.019$ times larger than from
our method.  We find below that the global mean velocity dispersion of
the combined dataset, normalized with our
\sig1\ values, is $\sigall$, whereas it should be unity, which is
consistent with either scheme for calculating the velocity
dispersions. Since the outcome of this investigation is that $M_v$ is
biased upwards, the bi-weight calculation would lead to a further
increase in the bias.

For each cluster a ``ringwise'' estimate of the virial radius
of the observed distribution is calculated (see \cite{global} for
details):
\begin{equation}
{1\over r_v}
={2\over \pi}\left(\sum_i w_i\right)^{-2} \sum_{i<j}{w_iw_j 
        {2\over{\pi(r_i+r_j)}} K(k_{ij})},
\label{eq:rv}
\end{equation}
where $k_{ij}^2=4R_iR_j/(R_i+R_j)^2$ and $K(k)$ is the complete
elliptic integral of the first kind in Legendre's notation
(\cite{nr}).  The angular extent of the sample is set by the observed
field size.  The redshift range is as found in the
velocity dispersion calculation.

The goal is that our data will include the entire virialized mass of
the cluster. Analytic models (\cite{gg}) and simulation data
(\cite{cole_lacey,zembrowski}) find that the virialized mass is
generally contained inside the surface where the mean interior density
is $200\rho_c$ at the redshift of the cluster. The mean interior
density within our measured $r_v$ is
\begin{equation}
{\overline{\rho}(r_v)\over{\rho_c(z)}} = {1\over{\rho_c(z)}} {{3
                M_v}\over{4\pi r_v^3}}.
\end{equation}
In terms of the measured dynamical parameters this is
\begin{equation}
{\overline{\rho}(r_v)\over{\rho_c(z)}} = 
        {{6\sigma_1^2}\over{H^2(z) r_v^2}},
\label{eq:rho}
\end{equation}
where the Hubble constant at $z$ is
$H^2(z)=H_0^2[\Omega_0(1+z)^3+\Omega_R(1+z)^2+\Omega_\Lambda]$
(\cite{ppc}). Normally we take $\Omega_0=0.2$, $\Omega_\Lambda=0$ and
$\Omega_R=1-\Omega_0$ for an open FRW model.  The mean density inside
$r_v$ serves as a check as to whether the radial extent of the cluster
is sufficiently sampled to determine a virial radius that includes
most of the virialized mass. Table~\ref{tab:r200} gives the values for
$r_v$, $\sigma_1$, and $\overline{\rho}(r_v)/\rho_c(z)$, in columns 3,
4 and 5, respectively. The sampling over-density
$\overline{\rho}(r_v)/\rho_c(z)$ varies substantially from cluster to
cluster, which mainly indicates the radial extent of our coverage of a
given cluster.  The observations used here extend beyond \r200 for
most clusters. The exceptions are MS1006$+$12, MS1008$-$12, and
MS1455$+$22 where the data extend to about 2/3 to 3/4 of \r200.

The cluster's virial mass to k-corrected Gunn $r$ luminosity,
$M_v/L$, and its errors are given in columns 6 and 7 of
Table~\ref{tab:r200}. The numbers are slightly different than those
found in
\cite{global} because these results are based on the finalized catalogues.

To calculate \r200, the radius where the mean interior density is
$200\rho_c(z)$, requires an assumption as to how the mass is
distributed.  One possibility is that the mass is strongly
concentrated to the center of the cluster such that all the mass
measured by $M_v$ is inside the measured $r_v$, in which case,
\begin{equation}
\rp200 = 
r_v \left[{{\overline{\rho}(r_v)}\over{200\rho_c(z)}}\right]^{1/3}.
\label{eq:rp200}
\end{equation}
This definition of $r_{200}$ is not particularly appropriate when
$\overline{\rho}/\rho_c(z)<200$, as it is for our best sampled
clusters, because the cluster galaxies, and presumably the mass as well,
extend beyond $r_{200}$. A superior \r200\ is calculated assuming
a density-radius model which is extrapolated to the desired
overdensity.  Simplicity, and a great deal of work on clusters of
galaxies and the properties of dark halos in general, suggests that
the first approximation for the density-radius relation is the
singular isothermal sphere, $\rho\propto r^{-2}$, in which case we
define,
\begin{equation}
\r200 =  r_v 
	\left[{{\overline{\rho}(r_v)}\over{200\rho_c(z)}}\right]^{1/2},
\end{equation}
or equivalently, 
\begin{equation}
\r200 =  {\sqrt{3}\sigma_1\over {10H(z)}}.
\label{eq:r200}
\end{equation}
Note that $\rp200\propto\sigma_1^{2/3}r_v^{1/3}$ and
$\r200\propto\sigma_1$.  That is, \r200\ is independent of the angular
extent of the sampling, in so far as the cluster's \sig1\ value has no
radial dependence, which for the shallow gradients of the velocity
dispersion profile found below is effectively true. At $z=\onethird$ a low
density, flat universe gives a value of \r200\ about 13\% larger than
in a open universe.

Values of $\rp200$ and \r200\ are given in columns 8 and 10 of
Table~\ref{tab:r200} in units of physical \hmpc\ for our cluster
sample. For our sample these two \r200\ estimates have very similar
values, so the results are not sensitive to the mass model. The
jacknife error estimates for $\rp200$ and \r200\ in \hmpc\ are given
in columns 9 and 11, respectively.  All of the following analysis will
use \r200\ as the scaling radius.

Two of the clusters, MS0906+11 and MS1358+62, are strong binaries, the
first having an ill determined velocity dispersion and the second
having a massive substructure on one side moving at 1000 \kms\
relative to the cluster center. These are dropped from all further use
in this paper.  The scaled velocity-radius data for the remaining
clusters are displayed in Figure~\ref{fig:cvt} plotted using the
brightest cluster galaxies (BCGs) as the projected center and the
weighted average redshift as the velocity center.  

\section{Background Subtraction}

All of the cluster properties presented in this paper are measured
using galaxies with redshifts.  The clusters typically extend over a
velocity range of $5000 \kmps$, which in redshift space causes
orbitting cluster galaxies to be intermingled with field galaxies
within a 50\hmpc\ distance interval around the cluster.  Background
and foreground galaxies that are projected into the redshift space of
the cluster must be statistically removed. This survey was
deliberately placed at moderate redshift to give a field sample over a
considerable redshift column from which the background density can be
accurately measured.  The background density is measured in the
normalized velocity-radius space of Figure~\ref{fig:cvt}.  It should
be noted that the sampling in Figure~\ref{fig:cvt} varies
substantially with both velocity and radius which is addressed with
redshift cutoffs and magnitude and position dependent weights
(\cite{yec}). The background and foreground are folded on top of one
another within the unbiased redshift sample range of the band limiting
filters which removes any linear gradients present in the redshift
distribution.

There are two approaches to background subtraction used in the
following calculations. For the surface density calculation we use the
straightforward average over the sky area observed. The velocity
dispersion calculation takes a more cautious approach to background
subtraction. A resolved background is derived in precisely the same
range of projected radii for which $\sigp(R)$ is calculated.  The
advantage of this approach is that the contamination of the cluster
velocity profile with various nearby groups of galaxies is subtracted
in the range of projected radii where they contribute to the
background. Furthermore, the $\sigp(R)$ calculated from
non-overlapping radial ranges are completely independent of each
other.

\section{The Average Density Profile}

The surface number density profile of the average cluster,
$\Sigma_N(R)$, is the projection over velocity onto the radius axis,
$R$, of the data of Figure~\ref{fig:cvt}.  To find $\Sigma_N(R)$, we
sum the statistical weights (which correct for nonuniform sampling,
\cite{yec}) for objects within 3\sig1\ of zero velocity,
divide by the bin area, and subtract the mean background.  We prefer
the BCGs as the projected centers (\cite{bird_bcg}), but repeat most
of the analysis using the peak of the smoothed density profile as the
centers, which gives a useful view of systematic effects in the
analysis.

The mean background volume density in velocity space,
$\overline{\rho}_b^\dagger$, is calculated by summing over the entire
range of $R/\r200$ using the background in the velocity range of 5 to
25 \sig1. We find that $\overline{\rho}_b^\dagger=0.0065\pm0.0008$, in
units of the normalized total numbers per \sig1 per $\r200^2$.
Figure~\ref{fig:back} shows the variation of the average background
density with velocity separation (in normalized units) from the
cluster. Between 3 and 6\sig1, the background calculated from the full
sample (left panel) is higher than expected for a high velocity tail
like a Gaussian. This excess is statistically significant at about the
3 standard deviation level, and results from various small groups that
happen to be behind the low redshift clusters.  The right panel of
Figure~\ref{fig:back} shows the background calculated after the 7
clusters with $z\le 0.27$ are dropped. The rapid decline of clustering with
redshift leads to a smoother distribution in redshift in the right
panel.  The mean background is calculated over the entire 5 to
25\sig1\ interval for the full sample, so a local excess makes little
difference.  As a sidelight we note that near the cluster center,
1-3\sig1, the velocity distribution is possibly somewhat ``squarer''
than a Gaussian distribution. That is, for a Gaussian the ratio of the
numbers between 1 and 2\sig1\ and 0 to 1\sig1\ is about 1/3, whereas
we have more than 1/2.  This is likely an indicator that the cluster
velocity ellipsoids are somewhat radially anisotropic
(\cite{dubinski,cer}).

To check the sensitivity of the derived
$\Sigma_N(R)$ to the assumed background level we repeat the
calculation using values for $\overline{\rho}_b^\dagger$ of 0.0048 and
0.0082, which are 2 standard deviations from the mean value. The only
part of the profile substantially affected is beyond \r200, with
alternate values being just within the 1 standard deviation range of
the estimated density at that point. That is, for our adopted bin
widths the statistical error due to sample size generally exceeds the
uncertainty in the background subtraction. The fitted profiles
described below have scale radii altered by 1.5\% for a 26\% variation
of the background. We conclude that the background subtraction is not
a dominant source of error for $\Sigma_N(R)$.

The background subtracted $\Sigma_N(R)$ are displayed in
Figure~\ref{fig:surf} and tabulated in Table~\ref{tab:surf}.  The
vertical normalization of $\Sigma_N(R)$ is a fixed, but arbitrary,
value.  In Table~\ref{tab:surf} the number density $\Sigma_0$ (column
2) is measured about the BCG, and $\Sigma_p$ (column 4) is measured
about the peak of the 45\arcsec\ Gaussian smoothed density map.
Within $0.1\r200$ of the center the density profile is strongly
dependent on the choice of center. Because there is usually no galaxy
at the point of peak density, there is a slight dip in the central
surface density when measured about the peak of the smoothed density
distribution. The choice of center makes little difference to the
measured $\Sigma_N(R)$ at radii beyond $0.1\r200$, approximately
100\hkpc, but it does affect fits to a specific model.  The point of
peak density generally has sufficiently large errors that it could be
consistent with being at the BCG, hence we always prefer the BCG
center for physical reasons (\cite{bird_bcg}).  Outside of $2\r200$
the surface density is comparable to the background but the data are
very sparse, which together cause the surface density at large radius
to be quite uncertain.  No data beyond $2\r200$ are used in the
velocity dispersion analysis.  The surface luminosity density profile
was also derived; but, being weighted to the brightest galaxies, it is
a noisier quantity than the number density. We find that the
luminosity density profile shape is essentially the same as the
density distribution.

The spatial galaxy number density as a function of radius, $\nu(r)$,
is needed to compare the galaxy number profile with the cluster mass
profile.  The relation between the volume density and the surface
density is,
\begin{equation}
\Sigma_N(R) = 2\int_R^\infty \nu(r) {r \over\sqrt{r^2-R^2}} \,dr.
\label{eq:surf}
\end{equation}
A statistically adequate, analytically convenient, model that describes these
data is the Hernquist (alternatively designated as an $\eta=2$) model
(\cite{hernquist,eta}).  We fit the data centered on the BCG
of Figure~\ref{fig:surf} to the projection of the volume density model,
\begin{equation}
\nu(r) = {A\over{r(r+a)^3}}.
\label{eq:nu}
\end{equation}
This procedure for finding a $\nu(r)$ consistent with the
$\Sigma_N(R)$ data is not unique, there being both the minor variation
of using one of several other entirely reasonable fitting functions,
or, adopting a completely different approach such as the non-parametric,
maximum likelihood method (\cite{mt}) modified to allow for background
subtraction.  The function $\nu(r)=Ar^{-1}(r+a)^{-2}$ (\cite{nfw}) is
also a statistically acceptable fit with a slightly increased
$\chi^2$.  Our measured $\Sigma_N(R)$ falls faster than this function
predicts at large $R$, but at these radii our measurement is
relatively uncertain.

Fitting the surface density data with Equation~\ref{eq:nu} finds that
$a=(\afit)\r200$ for the clusters centered on their BCGs.  The
reduced chi-squared statistic, $\chi^2/n_f$, is 0.74 for $n_f=17$
degrees of freedom, which is an acceptable fit.  Using the
$\Sigma_N(R)$ measured about the peak light gives a considerably
larger scale radius, $a=(0.97\pm0.10)\r200$, with
$\chi^2/n_f=1.01$. This larger value of $a$ is at least in part due to
a blurring of the density profile by statistical errors in measuring
the position of peak number density. Although we believe it is a less
appropriate choice for the cluster centers for these data, we carry it
along in the analysis as an indicator of systematic errors.

\section{The Velocity Dispersion Profile}

The measurement of $\sigp(R)$ is a second moment of the velocity
distribution about the local mean velocity at projected radius $R$,
although the choice of central moment makes essentially no difference
to the result.  The projected velocity dispersion at $R$, $\sigp(R)$,
is calculated using an iterative clipping of the high velocity
dispersion tails with a technique similar to that used for the RMS
velocity dispersion in
\cite{global}, where $\sigma_1$ is calculated from the background
subtracted data within $n_c\sigma$ of zero velocity, where $n_c=3$.
The iteration begins with $\sigp(R)=1$ in the \sig1\ normalized
velocities.  For the next four iterations the $\sigp(R)$ from the
previous iteration is used for clipping.  For a 10\% change in the
clipping level, $n_c$, the resulting velocity dispersion changes about
1\% at large radii.  The error flags are calculated from a full
bootstrap resampling of both the cluster and field data.

Rather than radially binning the data, we use a moving average, which
includes all galaxies, both cluster and field. Table~\ref{tab:sig}
gives the results with a 51 point moving average which are plotted in
Figure~\ref{fig:sigfit} at intervals of 0.1 in
\r200. These points are always independent for 51 point averaging.  
Averages using 31 to 201 points give consistent results, the 31 point
result being quite noisy, and the 201 point beginning to smooth the
velocity dispersion gradient.  The results of a 101 point moving
average are shown in Figure~\ref{fig:sig}. For 101 point averaging the
error flags are not independent outside of $0.4\r200$.

The random errors of the derived $\sigp(R)$ of the ensemble are
dominated by cluster to cluster fluctuations. Within an individual
cluster (for instance A2390) the random errors are substantially
smaller, but there are large differences in velocity dispersion
profiles from cluster to cluster (\cite{hartog}) which we attribute
to projection effects and substructure.  At large radii, dynamical
measurements from a single cluster likely do not give reliable
enclosed masses because of the complications of infall, triaxiality
and substructure.  The idea of creating the ensemble cluster is of
course to diminish these systematic variations to the level where an
effectively spherically symmetric average profile emerges.

To derive a $\sigma_r(r)$ which is consistent with the observed
$\sigp(R)$ we use the same approach as we used for the density
profile. That is, we choose a reasonable functional form, then adjust 2
parameters, a scale length and a normalization, until its calculated
projection,
\begin{equation}
\sigp^2(R)\Sigma_N(R) = \int_R^\infty \nu(r)\sigma_r^2
	(1-\beta{R^2\over r^2}) {r\over\sqrt{r^2-R^2}} \,dr.
\label{eq:sigp}
\end{equation}
minimizes the $\chi^2$.  In this equation $\beta= 1
-\sigma_\theta^2/\sigma_r^2$ is the velocity anisotropy parameter.
For the purposes of this paper $\beta$ will be taken to be a constant,
although we do use a range of values.

The adopted $\sigma_r(r)$ function should be
finite at the origin, at large $r$ it should tend to a Keplerian
$r^{-1/2}$ for a convergent mass distribution, and it should be a
simple smooth function in between. It is important that this function
not assume that the galaxy populations are self-consistent with the
mass density of the potential. We adopt the simple form,
\begin{equation}
\sigma_r^2(r)= {B\over{b+r}},
\label{eq:sigfit}
\end{equation}
for the radial velocity dispersion.  This function, with
B=$\onequarter$ and assuming $b=a$, is the solution of the
Jeans Equation (Equation~\ref{eq:jeans} below) for $\beta=\onehalf$
and $A=1/(2\pi)$ in our assumed $\nu(r)$.  The scale lengths
$b$ and $a$ are fitted separately, usually finding that $b$ 
is larger than $a$.

The nonlinear fitting procedure requires that the errors be symmetric
about the observed value, so we take the larger of the upper and lower
error range at each point.  The fit is done assuming
$\beta=0,0.25,0.5$ and $0.75$.  The projected fit and the 51 point
data are shown in Figure~\ref{fig:sigfit}. The $\chi^2/n_f$ of the two
parameter fit, an amplitude and length scale, is $0.48$ for all
$\beta$ values.

\section{Mass-to-Light Profiles}

To compare the relative distribution of mass and light we integrate
our fitted $\nu(r)$ to give $L(r)$, the ``light profile'' (actually a
number density profile), which is converted to a mass profile using a
normalized global $M_v/L$, calculated in the same manner as for the
individual clusters. Then we compare the mass predicted by the
light-traces-mass assumption to the $M(r)$ derived from the fitted
$\sigma_r(r)$ and $\nu(r)$ using Jeans Equation,
\begin{equation}
M(r) = -{\sigma_r^2r\over G}
        \left[{{d \ln{\sigma_r^2}}\over{d\ln{r}}} +
        {{d\ln{\nu}}\over{{d\ln{r}}}} +2\beta\right].
\label{eq:jeans}
\end{equation}
We will usually refer to the mass derived from this equation
as $M_{SHD}(r)$, the stellar hydrodynamical mass.

It must be borne in mind that Equation~\ref{eq:jeans} is a moment of
the Collisionless Boltzmann Equation, and will not necessarily give
a physical result for all $\beta$, being particularly susceptible to
failure for very radial orbits (\cite{bt_gd,hernquist}).
In addition, radial models are usually dynamically unstable to the
radial orbit instability (\cite{bgh86,pp87}). However, the galaxies
are only tracers, so in principle they could be on much more radial
orbits than the underlying dark matter. On the other hand, it seems
likely that galaxies orbiting past the center may be tidally
destroyed or simply well mixed so that the galaxy velocity ellipsoid
near the center may become less radial than the dark matter.

\subsection{Dynamical Modeling Details}

Equation~\ref{eq:jeans} is integrated with the boundary condition that
$\sigma_r\rightarrow 0$ as $r\rightarrow \infty$. Alternative boundary
conditions are that the velocity dispersion goes to zero at the
turnaround radius (\cite{gg}), or, that there is an outer region of
nearly radial infall which makes a transition to virialized motion
(\cite{hartog}). N-body simulations (\eg\
\cite{cer,cole_lacey,zembrowski}) show that on the average, the
approximation that the virialized motion continues to \r200\ is 
a good one.  For constant $\beta$ Equation~\ref{eq:jeans} has a
formal solution,
\begin{equation}
\sigma_r^2(r) = {{\int_r^\infty G M(x)\nu(x)x^{(2\beta-2)}\,dx}\over
	{\nu(r) r^{2\beta}}}.
\label{eq:solution}
\end{equation}
Satisfying the Jeans equation is a necessary condition for an
equilibrium to exist, but it is not sufficient and is certainly not a
guarantee that a spherical distribution is dynamically stable.  Models
with a constant $\beta\gtrsim\twothirds$ are unphysical, because the
implied mass is negative in the inner regions (very radial orbits
lead to $\nu(r)\propto r^{-2}$, not the $r^{-1}$ here). This, along
with the results of numerical simulations, leads to our preference for
$0\le\beta\le\onehalf$. The model lines plotted on
Figure~\ref{fig:sig} are for $\beta=0,\onehalf$ and 1, assuming
$M(r)\propto L(r)$ in Eq.~\ref{eq:solution}, using the fitted $\nu(r)$
from Eq.~\ref{eq:nu}, and normalizing the projected RMS velocity
dispersion over the observed radial range to the observed value.

The numerical complication in deriving $M(r)$ from
Equation~\ref{eq:jeans} is that the logarithmic gradients
$d\ln{\sigma_r^2}/d\ln{r}$ and $d\ln{\nu}/d\ln{r}$ must be calculated.
Numerically evaluating these gradients from the data requires some
form of smoothing to be stable, which our fits to smooth functions
effectively accomplish.  We will carry the errors of the projected
\sigp\ data through the analysis to help give an assessment of the
result.

\subsection{Velocity Ellipsoid Anisotropy}

The radial dependence of the shape of the velocity ellipsoid, as
defined by the value of $\beta$, is not empirically known for this
sample, although this does not introduce the same uncertainty as it
does for finding central dark masses.  As the velocity ellipsoid
becomes more and more radial at large radii, the mass implied for a
measured \sigp\ increases.  The inferred mass has nearly a full order
of magnitude uncertainty when a velocity dispersion is available at
only a single point with no knowledge of the shape of the velocity
ellipsoid (\cite{merritt,rt84,tw84}).  Because we have measured the
velocity dispersion over a wide radial range the uncertainty due to
velocity dispersion anisotropy becomes quite small, as shown below.

N-body simulations of objects that form via collapse find that
$\beta\ge0$ (\cite{rt84}), although it is clear now that ``vacuum''
boundary conditions give far more velocity anisotropy than is
realistic in a cosmological setting (\cite{dubinski}) where
substantial external torques provide angular momentum to large radius
orbits. Furthermore, strongly radial models are unstable
(\cite{bgh86,pp87}), although galaxies, being nearly massless within a
cluster, could have a different velocity ellipsoid than the dark
matter.  For the nearly power law hierarchical clustering that
characterizes the growth of clusters in model universes one expects
that the velocity ellipsoid will have only a weak radial dependence,
because cluster buildup is self-similar. This is borne out in n-body
simulations (\cite{dubinski,cer,zembrowski}) which find
$0\le\beta\le0.5$ over the radial range of interest for a wide range
of $\Omega_0$ and clustering models. The depletion of HI in Coma
spirals suggests (\cite{pg84}) that their velocity ellipsoid is not
strongly radial, and is consistent with being isotropic.

\subsection{The Average Mass-to-Light Ratio within \r200 }

A test of the accuracy of the virial mass as an estimator of the total
mass virialized in the cluster is whether the product of the virial
mass-to-light ratio computed from our average cluster and the
integrated interior luminosity, $L(r)$, is equal to the stellar
hydrodynamical mass, $M_{SHD}(r)$, derived from
Eq.~\ref{eq:jeans}. That is, we derive the virial mass bias,
$b_{Mv}(r)$, from
\begin{equation}
b_{Mv}(r) = {{M_{SHD}(r)}\over{L(r)}}  {\tilde{L} \over\tilde{M}_v},
\label{eq:mlbias}
\end{equation}
where $L(r)$ is simply the volume integral of $\nu(r)$.  The
normalization of $L(r)$, $\tilde{L}$, is arbitrary since it cancels in
Eq.~\ref{eq:mlbias}.  The normalizing virial mass, $\tilde{M}_v$, is
derived for the ensemble cluster in the same way as for the individual
clusters.  Two aspects of $b_{Mv}(r)$ are of primary interest.  First,
$b_{Mv}(\r200)$ should be unity if the $\tilde{M}_v/\tilde{L}$
calculated from the average cluster is an unbiased estimator for all
material inside
\r200, the radius expected to contain most of the virialized mass.
Second, the radial gradient of $b_{Mv}(r)$ measures to what degree
light traces mass in the average cluster. The first issue is the one
of most relevance for the problem of $\Omega$ estimation.  It should
be noted that whatever the value of $b_{Mv}(r)$ turns out to be here,
it is a quantity that is meant to give the systematic error in
measuring cluster mass using red selected galaxies in rich, reasonably
relaxed clusters. It would be inappropriate to apply the same factor
to merging clusters or a galaxy sample selected to emphasize high star
formation rates, for example.

The galaxy number density profile, which we will call the light
profile, is normalized to a mass using a mass-to-light ratio
calculated in the same manner as done for our clusters individually.
There are two complications in this calculation which need to be
examined.  First, the clusters are not all sampled to the same
fraction of
\r200. Second, both the $r_v$ and the total light are calculated
without any allowance for background contamination.  To investigate
these effects we calculate $\tilde{M}_v/\tilde{L}$ using the mean
profile, both with and without background subtraction, limiting the
data at various $R_{max}/\r200$. The results are shown in
Figure~\ref{fig:mlrad}.

Remarkably, the ratio $\tilde{M}_v/\tilde{L}$ is almost entirely
independent of background contamination and sampling radius, once
$R_{max}/\r200\gtrsim 0.5$.  The lower line on the figure is made
using the surface density without background subtraction, which causes
$\tilde{r}_v$, hence $\tilde{M}_v$, to be overestimated, but the
summed luminosity is also overestimated. The pleasing result for this
sample is that the two overestimates almost exactly cancel out so that
$\tilde{M}_v/\tilde{L}$ calculated in the usual manner has no
sensitivity to background. Furthermore, $\tilde{M}_v/\tilde{L}$ has
very little dependence on the sky coverage of the cluster, provided
that the data extend beyond $0.5\r200$. Figure~\ref{fig:mlrad} shows
that the directly calculated average $\tilde{M}_v/\tilde{L}$ is high
in the inner regions, but then drops to a very stable value at large
radius.  The normalizing virial mass is
$\tilde{M}_v=3\tilde{\sigma}_1^2
\tilde{r}_v$, where $G=1$ and the parameters are the measured
$\tilde{\sigma}_1$ and $\tilde{r}_v$ for the combined cluster data.
We use the quantities for the full sample, with background
subtraction, although this choice makes little difference to the
results.  With this normalization, $b_{Mv}(\r200)$ should be unity if
$M_v$ is unbiased.

\subsubsection{Virial Mass Bias}

The internal $M_v/L$ bias, $b_{Mv}(r)$, from Eq.~\ref{eq:mlbias} is
displayed as a function of radius in Figure~\ref{fig:mol} for
$\beta=0,0.25,0.5$ and $0.75$.  The bias would be
unity everywhere if the product of $\tilde{M}_v/\tilde{L}$ and $L(r)$
was equal to the stellar hydrodynamical mass.  Note that if
$\beta\gtrsim 2/3$ the implied mass is negative at small radius.

The primary goal of this paper is to correct the virial mass-to-light
ratio to the field $M/L$ value.  The average virial mass bias for the
bulk of the virialized system estimates the systematic offset in the
virial mass, and is reasonably estimated as the value of $b_{Mv}$
evaluated at \r200, displayed in Figure~\ref{fig:mvb} as a function of
the assumed $\beta$.  Because the gradient is small the exact radius
used does not make a big difference. The error in $b_{Mv}(\r200)$ is
calculated from the error in the fit to the $B$ coefficient in
$\sigma_r(r)$, the error in the other quantities being negligible in
comparison.  The 3 symbols are for absolute magnitude cutoffs of
$M_r^k=$-18.5, -19.0, and -19.5 mag, respectively, of the entire
calculation. The points offset in $\beta$, with the dashed error
flags, use the peak density as the centers and are always poorer fits,
hence have larger errors. These points are included to illustrate the
small systematic differences. All the results are statistically
identical.

We find $M_v$, as normally calculated, is always an overestimate of
the mass, whose value is weakly dependent on the assumed $\beta$
value.  For this sample the bias is minimal (\ie\ $b_{Mv}$ closest to
unity) near $\beta<\onehalf$ for which it is equal to $\mvadj$.  All
$\beta$ give statistically consistent bias values, although there is
clear trend of decreasing $b_{Mv}$ with increasing $\beta$.

\subsubsection{The Mass-to-Light Ratio Gradient}

The inferred radial gradients of $b_{Mv}$ for different $\beta$ are
shown in Figure~\ref{fig:mvg}. The gradient is calculated from the
slope of the lines in Figure~\ref{fig:mol}.  Only for very radial
orbits, $\beta\ge \twothirds$, is there even a mildly significant
rising mass-to-light ratio with radius, and those are the models with
negative central mass.  The gradients are sufficiently small that for
a wide range of $\beta$ they are consistent with there being no
gradient of $M/L$. In no case are the gradients sufficiently large
that there is any possibility that the cluster has a dark matter halo
that is significantly more extended than the visible galaxies.

\subsection{A Model Calculation of the Virial Mass Bias}

We have found that the virial mass, as we have calculated from the
data, is about a 20\% overestimate of the true mass.  The origin of
this factor needs to be quantitatively understood.  We believe that
this bias is simply the approximation that the surface pressure term
in the virial equation is zero.  For our fitted functional forms for
the density profile and the velocity dispersion profile, (Equations
\ref{eq:nu} and \ref{eq:sigfit}, respectively), we can provide an
estimate of the expected ``truncation bias''.

Truncating a density profile of the form Eq.~\ref{eq:nu},
$\rho(r)=Ma(2\pi r)^{-1}(r+a)^{-3}$, will cause the true mass
contained within that radius to be overestimated, as quantified
below. The potential energy scalar, $W(x)$, of the density
distribution truncated at $x$ is,
\begin{equation}
4 \pi \int_0^x \rho M(r) r \, dr = 
{\frac {GM^2{x}^{3}\left (x+4a\right )}{6a\,\left (x+a\right )^{4}}},
\label{eq:w}
\end{equation}
Calculating the virial radius as $r_v=M(x)^2/W(x)$ we find
that
\begin{equation}
r_v(x)= {\frac {6\,xa}{x+4a}}.
\label{eq:rg}
\end{equation}
The estimated virial radius is $6a$ if the distribution is sampled to
infinity distribution, whereas for our combined clusters we have
measured the virial radius to be $3.9a$ (using the $a$ found from the
fit of the distribution with background subtraction). This is
consistent, within the errors, with our truncation radius $x\simeq4a$
and Eq.~\ref{eq:rg} for the combined sample. The clusters as
individuals all have truncation radii less than or equal to this.

If the galaxy distribution follows the mass density the mass contained
within $x$ is estimated from the ``classical'' virial theorem relating
the kinetic energy, $T(x)$, $W(x)$, and the surface term, $3PV(x)$,
through $2T(x)+W(x)=3PV(x)$, as $M_v(<x)=2G^{-1}T(x)r_v(x)/M(x)$. The
usual assumption for dynamical systems is that $3PV(x)$ is evaluated
as $x\rightarrow\infty$, where it is zero. This is not correct in
detail for clusters of galaxies, where this term is a significant
correction.

For the particular case of $\beta=\onehalf$, it was noted above that
the solution of Jeans Equation is $\sigma_r^2(r) = GM/[4(a+r)]$.
Hence the truncated kinetic energy, $T(x)$, is,
\begin{equation}
2\pi(3-2\beta)\int_0^x \rho(r) \sigma_r^2(r) r^2\,dr =
{\frac {GM{x}^{2}\left (x+3a\right )}{12a\,\left (x+a\right )^{3}}}.
\label{eq:tvir}
\end{equation}
For the case of the mass being distributed like the galaxies, the
ratio of the estimated mass contained within $x$ to the true mass
contained is,
\begin{equation}
{{M_v(<x)}\over{M(x)}} = {\frac {\left (x+3a\right )\left
(x+a\right )}{x\left (x+4a\right )}}.
\label{eq:vmx}
\end{equation}
Note that this expression could also be derived from $2T+W=3PV$, as
discussed in \cite{global}.  At any finite radius this estimated mass
is greater than the true global value.  At $x=a$, $2a$, $3a$ and $4a$
it evaluates to 8/5, 5/4, 8/7 and 35/32, respectively. That is, the
mass estimated from the averaged distribution is expected to
overestimate the true value by about 10 to 20\%. Clusters that are not
sampled to large radii will have a larger truncation bias but smaller
background contamination in the projected quantities. The data in most
of our clusters extend to 1 to 2 \r200, roughly consistent with the
measured $b_{Mv}$.

\section{The Radial Gradient of Galaxy Populations}

A crucial step in the derivation of $\Omega$ from cluster $M/L$ values
is to correct for any differential luminosity evolution between the
cluster and field galaxies.  The luminosities of galaxies per unit
total mass can, in principle, either increase or decrease depending on
what happens when galaxies fall into a cluster.  The empirical
evidence is that once galaxies have spent some time in rich, high
X-ray luminosity clusters like those in our sample, star formation
largely ceases. However, there are at least two routes to this end
point for a gas rich disk galaxy. Galaxies entering the cluster could
have their gas largely stripped away, or, a burst of star formation
upon cluster entry could boost luminosity prior to depleting the
gas. In either of these cases, lowering the star formation rate leads
to a decrease in a galaxy's luminosity by an amount that depends on
its entire star formation history. Galaxy merging always decreases
numbers, but can lead to luminosity increases if accompanied by star
formation.  All these possibilities are testable in one way or
another, and some substantial constraints on differential evolution
can be imposed using the data at hand.

It is well known that the fraction of cluster galaxies with blue
colors is a strongly increasing function of redshift, the
Butcher-Oemler effect (\cite{bo,cnoc_bo}).  It should be noted that
cluster galaxies are, on the average, never bluer than field galaxies,
implying that, on the average, star formation always declines into a
cluster.  If the infall of field galaxies into clusters induces a
temporary increase star formation and hence increases cluster galaxy
luminosities relative to the field, then it should be apparent as a
change in the color distribution (and the strength of the [OII]
lines).  This is not seen in our data (\eg\ \cite{a2390}). On the
basis of these considerations we expect that cluster galaxies are of
intrinsically lower luminosities than the field galaxies.

Our sample of galaxies extends from the cluster center to the distant
field, allowing population gradients to be examined as a function of a
redshift space radial variable, $s$.  A convenient definition of such
a variable is based on our normalized co-ordinates, where velocity
differences are scaled to \sig1\ and projected radial co-ordinates are
scaled to \r200, so that the dimensionless redshift space separation
from the center is $s^2=R^2/\r200^2+(\Delta v)^2/\sigma_1^2$, where
$R$ is the projected radius from the center and $\Delta v$ is the
rest-frame velocity difference from the average redshift.  Note that
at large $s$ the galaxies come exclusively from the redshift
direction.  Figure~\ref{fig:czr} shows the color-radius relation of
the full sample.  All galaxies with k-corrected absolute $r$
magnitude, $M_r^k$, brighter than $-19$ mag are plotted.  The colors
are corrected to an empirically normalized
redshift-independent color,
\begin{equation}
(g-r)_z = (g-r) [1.2+2.33(z-0.3)]^{-1}
\label{eq:gr}
\end{equation}
(\cite{global}).  We note that there is no color gradient in the
field, but one appears at $s\simeq2$, where we expect galaxies to
enter the virialized region of the cluster.  There is no evidence in
the colors for excess star formation at the edge of the cluster;
indeed, there are no cluster galaxies systematically bluer than field
galaxies. Hence, the distribution of galaxy color with radius
(Fig.~\ref{fig:czr}) is consistent with infalling galaxies terminating
their star formation with little or no ``starbursting'' once they
enter the cluster, as detailed analysis of the CNOC data for A2390 has
demonstrated (\cite{a2390}).

A sample extending to a larger projected radius is needed to directly
demonstrate kinematic evidence for infall of galaxies into the
clusters.  However, given the gravitational mass of the cluster, the
turnaround radius is straightforwardly derived (\cite{gg,kg}) such
that field galaxies must be falling into the cluster out to the
turnaround radius, approximately $4.6\r200$ for the model profile
here. To a significant degree the cluster must be composed of former
field galaxies. In such an infall scenario the cluster galaxies
statistically share the star formation history of the field, prior to
cluster entry.

The only gross difference between clusters and field is that
clusters contain an unusual population of luminous central galaxies.
The BCGs only make a difference at small $s$, and generally
contribute well less than 10\% of the light.  In spite of the substantial
radial color gradient of Figure~\ref{fig:czr} there is very little
change in the $r$ band luminosity distribution of
Figure~\ref{fig:lzr}. Measurements of the surface brightness evolution of
both disks and bulges arrives at the conclusion that the amount
of differential evolution between the field and the cluster luminosities
is relatively small (\cite{schade_e,schade_d}).

There is no significant difference in the field galaxy luminosities or
colors at any radius beyond $2\r200$, as shown in
Figure~\ref{fig:lzr}. That is, the field galaxies near the cluster
appear to be indistinguishable from those far away. A generic
prediction of ``natural bias'' models of cluster galaxy formation is
that the galaxies that eventually fall into the cluster should form
earlier because the overdensity of the incipient cluster speeds up the
local time scale relative to the mean universe. Possibly this leads to
a higher rate of conversion of gas to stars, causing cluster galaxies
to be more luminous than their counterparts in the field.  The color
gradient appears at about $2\r200$, approximately where one expects
cluster X-ray gas to be encountered.

Any fading or brightening of galaxies as they fall into the cluster
cannot be large.  Without the BCGs we measure a mean $M_r^k$ in the
average cluster, $r\le\r200$, which is statistically identical to the
field.  However, this averages together blue and red galaxies, which
contribute differently to the field and cluster, masking a real
fading.  Splitting at $(g-r)_z=0.7$ (Eq.~\ref{eq:gr}) finds that both
the blue and red galaxies show a decline in their mean brightness,
above our limit, of $0.11$ mag ($\pm0.05$ for red, $\pm0.07$ for blue)
between field and cluster.  Splitting the sample into a low and high
redshift sample shows that the fading between red galaxies in cluster
and field increases to about 0.2 mag at low redshift. A full
luminosity function fit (H.~Lin, private communication) finds a 0.2
mag field to cluster fading, but with large errors correlated with the
faint end slope.  In any case there is no evidence within these data
for an excess in stellar luminosity per unit dark matter in cluster
galaxies over field galaxies.  The corrected $r$ band mass-to-light
ratio in the field and these clusters does not vary much simply
because there is so much ``old'' red light which is not altered during
cluster infall.

\section{Corrections and Error Analysis}

The $\Omega_0$ estimate (\cite{global}) from cluster virial masses and
the conversion of the cluster luminosity (or numbers of galaxies) to
an equivalent co-moving volume in the field depends critically on the
accuracy of the virial mass and the average luminosity of field and
cluster being identical or having a measurable differential evolution.
The expression for $\Omega$ that we used as our best estimator is
$\Omega=(\langle M_v/L \rangle)/(\rho_c/j)$.  The field luminosity was
estimated to be $\rho_c/j=\mlclose h \msun/\lsun$ where the errors
were determined from a Jacknife analysis.  The value and random errors
of $\langle M_v/L \rangle$ were estimated as $\mlobs
h\msun/\lsun$. From these results we concluded that the virial mass
estimator gives $\Omega_0=\omlz$. However, this result needs to be
corrected for errors in $M_v/L$ relative to the field.

We find that the virial mass needs to be multiplied by
$b_{Mv}(\r200)=\mvadj$ to give the SHD mass at \r200. The correction
is a consequence of measuring the virial mass from a truncated
distribution and using data beyond \r200.  The values of $b_{Mv}$ for
$-1\le\beta\le 3/4$ are all within the errors of this value.  We also
find a significant mean luminosity differential between field and
cluster which requires that the cluster luminosity be boosted by
$\ladj$ mag to give the field value. Together these corrections reduce
$M_v/L$ to a field $M/L=\mlcor h \msun/\lsun$ and hence
$\Omega_0=\omzeroc$.  The errors in the various quantities are given
in Table~\ref{tab:err}, along with their quadrature and linear
sum. The main errors in our result are simply the statistics of the
luminosity density and the average cluster $M_v/L$ values. The
systematic errors are estimated from the extreme versions of the
analysis that we find for alternative $\beta$ values, choice of
center, and color and redshift samples of the galaxies.  To obtain an
$\Omega_0$ value accurate to about 10\% would require a sample
approximately four times larger.

The total random error, 30\%, is estimated from the quadrature sum of
the individual errors. Some of the component errors could be partially
correlated which would increase the total error somewhat.  The worst
case possibility is if all the errors are completely correlated, in
which case the error is the linear sum of the errors, 73\%. This must
be a substantial overestimate because many of the errors come from
completely uncorrelated sources.

The systematic errors in Table~\ref{tab:err} are harder to
estimate. Other choices of density centers and $\beta$ values give
answers that allow error estimation for these parameter
variations. The least satisfactory situation is the estimate of fading
between cluster and field.  Our observations contain no evidence for
excess star formation between cluster and field, so we feel that an
increase in the average light can be ruled out. Similarly, fading of
much more than 0.5 mag in our $r$ band light would lead to such gross
differences between cluster and field that it would be immediately
visible. Furthermore, for such a large fading to occur would likely
require a stellar population weighted towards short lived luminous
stars. To constrain further the differential luminosity evolution
requires more detailed observations of the individual galaxies.

\section{Discussion and Conclusions}

The overall goal of the CNOC survey is to use clusters of galaxies to
derive a value of $\Omega_0$ with a well determined error budget with
particular emphasis on eliminating systematic errors.  The major
innovations of our analysis are that it is completely self-contained,
with most assumptions being testable, and that the error estimates are
derived from the data themselves.  The dominant source of error is
random cluster-to-cluster variations due to projection and substructure,
rather than the internal error from individual clusters.  The global
mass-to-light ratio (in our photometric system) of our sample clusters
is constant within our typical errors of 20-30\% at a value of $\mlobs
h\msun/\lsun$. Over the same redshift range we measure the closure
value, $\rho_c/j$ to be $\mlclose h\msun/\lsun$ (\cite{global}). We
have shown that at
\r200\ that the ratio of the dynamical mass
to the cluster light, $b_{Mv}(\r200) =\mvadj$, with essentially no
dependence on assumptions about velocity anisotropy.  After allowing
for the fading of cluster galaxies relative to the field, $\ladj$ mag,
and the bias in estimating $M_v$, we estimate that $M/L$ field is
$\mlcor h \msun/\lsun$, implying $\Omega_0=\omzeroc$.  The objectively
evaluated random errors and estimated systematic errors are given in
Table~\ref{tab:err}.

A strength of these data is that they cover the entire virialized
cluster, extending out to where the overdensities are low, about
$15\rho_c(z)$, and hence reach to sufficiently large radii that the
cluster $M/L$ should represent the $M/L$ of the infalling material.
If segregation of dark matter mass and galaxies somehow occurs during
infall, prior to to virialization in the cluster, then that could
produce an artificially low $M/L$.  We find that the galaxy population
outside the cluster has no measurable gradients. A benefit of this
redshift range is that the galaxy populations in clusters more
strongly resemble the field than they do at low redshift, tightening
the $M/L$ argument.

Our $\Omega_0$ estimate is similar to the Least Action Principle
result of $\Omega_0=0.17\pm0.10$ (\cite{spt}) which probes larger
scales which have much smaller density contrast.  Our result is
inconsistent with an $\Omega_0$ of unity in any ``cold'' matter (less
than about 1000\kms) that falls into clusters.  Furthermore the result
is well below the $b=1$ interpretation of the large scale streaming
velocity results (\cite{dekel,sw,iras,potiras}). In particular our
result in combination with the Cosmic Virial Theorem estimates of
$\Omega_0/b\simeq0.2$ (\cite{dp,bean}) implying that these galaxies
are relatively unbiased mass tracers.  Because clusters assemble their
mass from regions approximately 20\hmpc\ across, the difference
between the streaming results and our determination is
substantial. That is, there is little room for extra matter which
supports density perturbations on larger scales but does not enter
clusters over our range of observed redshifts.

The $\Omega_0$ we derive is a reasonably comfortable 3 standard
deviations above the upper end of the fraction universe in baryons,
$\Omega_b$, (\cite{walker,copi}), as the X-ray mass measurements also
indicate (\cite{wnef,wf}). The X-ray analyses for the ratio of the
X-ray gas mass, $M_x$, to the total mass, $M_t$, find $M_x/M_t\simeq
0.09$ (approximately adjusted to $h=1$, but see \cite{wf}). In
combination with our results for $\Omega_0$ this line of reasoning
gives $\Omega_b =0.02h^{3/2} (0.2/\Omega_0) (0.1/(M_x/M_t))$ with an
error of about 35\%. Consequently cluster baryons are consistent with
other BBN indicators for $h\simeq0.7$, although with a fairly large
error range.

The fact that the galaxy numbers and the cluster mass are distributed
in an identical manner rules out any significant velocity bias between
the cluster galaxies and the dark matter.  That is, a small (say 20\%)
decrease in the velocity dispersion of cluster galaxies relative
to the dark matter would lead to a substantial segregation of the
cluster galaxies relative to the cluster mass
(\cite{wr,cd,kw,vbias}).  The conclusion to be drawn is that the
galaxy tracers selected in $\Omega=1$ n-body simulations do not form
like the galaxies in the real universe. The implications of this are
unclear. It is possible to identify relatively unbiased galaxy tracers
in n-body simulations (\cite{vbias}) so the problem may be within
the simulations. A more interesting possibility is that the
understanding of galaxy formation and mass clustering that is
incorporated into the simulations has some substantial deficiencies.

There are several improvements in this analysis that could be made
with more data, mainly at large radii.  More data at large radii would
reduce the variance resulting from substructure and sample all the
clusters to similar \r200, diminishing the complications of the
$\langle M_v/L \rangle$ bias correction.  With data extending beyond
$5\r200$ one enters the infall regime and expects to see the
``compression effect'' in the redshift space density contours of the
ensemble cluster (\cite{kaiser,regos_geller}), which would empirically
demonstrate that infall is occurring, and would give a measurement of
$\Omega^{0.6}/b$ in the nearly linear regime.  With many more
velocities in the cluster bodies the shape of the velocity ellipsoid
can be directly measured, rather than taken from n-body simulations as
was done here.

\acknowledgments
We thank the many members of CNOC who helped us obtain and reduce
these data. Comments from the referee, Doug Richstone, lead to
substantial improvements in the presentation.  We thank CFHT for the
technical support which made these observations feasible.  NSERC
provided financial support for RGC and HKCY.

\clearpage
\begin{table}[h]
\caption{Dynamical Parameters of the CNOC Clusters\label{tab:r200}}
\begin{tabular}{rrrrrrrrrrr}
\tableline
Name & $z$ &$r_v^\ast$ & $\sigma_1$ & $\overline{\rho}(r_v)/\rho_c(z)$ &
	$M_v/L$ & $\epsilon_{M/L}$ 
	& $\rp200$ & $\epsilon_{200}^\prime$ & \r200\ & $\epsilon_{200}$ \\
 & & \hmpc\ & $\kmps$ &  &\multicolumn{2}{c}{$h^{-1} \msun/\lsun$}
	 &\multicolumn{4}{c}{\hmpc} \\
\tableline
      A2390& 0.2280& 3.154& 1095&   46& 337&  54& 1.93& 0.12& 1.51& 0.08\\
MS0016$+$16& 0.5465& 1.639& 1243&  130& 260&  79& 1.42& 0.16& 1.32& 0.14\\
MS0302$+$16& 0.4245& 0.877&  656&  152& 260& 110& 0.80& 0.14& 0.77& 0.11\\
MS0440$+$02& 0.1965& 1.843&  611&   44& 383& 110& 1.12& 0.13& 0.87& 0.09\\
MS0451$+$02& 0.2011& 2.064&  979&   90& 435&  80& 1.58& 0.13& 1.38& 0.09\\
MS0451$-$03& 0.5391& 1.289& 1354&  252& 383& 110& 1.39& 0.14& 1.45& 0.11\\
MS0839$+$29& 0.1930& 0.805&  788&  389& 387& 147& 1.00& 0.19& 1.12& 0.16\\
$^{\dag}$MS0906$+$11& 0.1706& 0.790& 1834& 2282&1041& 238& 1.78& 0.11& 2.67& 0.17\\
MS1006$+$12& 0.2604& 0.890&  912&  377& 338& 117& 1.10& 0.13& 1.22& 0.15\\
MS1008$-$12& 0.3063& 0.893& 1059&  466& 301&  82& 1.18& 0.12& 1.36& 0.14\\
MS1224$+$20& 0.3255& 0.994&  798&  207& 330& 135& 1.00& 0.13& 1.01& 0.12\\
MS1231$+$15& 0.2353& 1.404&  662&   83& 235&  50& 1.05& 0.12& 0.91& 0.09\\
$^{\ddag}$MS1358$+$62& 0.3290& 2.393&  910&   46& 229&  30& 1.47& 0.10& 1.15& 0.07\\
MS1455$+$22& 0.2568& 1.027& 1169&  468& 810& 249& 1.36& 0.14& 1.57& 0.19\\
MS1512$+$36& 0.3727& 1.803&  697&   44& 413& 185& 1.09& 0.20& 0.85& 0.12\\
MS1621$+$26& 0.4275& 2.200&  833&   39& 201&  39& 1.27& 0.10& 0.97& 0.07\\

\end{tabular}
\smallskip

$^\ast$ Very dependent on the angular size of the region sampled

$\dag$ Strong binary with erroneous velocity dispersion not included in average

$\ddag$ Strong binary not included in average

\end{table}

\begin{table}
\caption{Ensemble Surface Density \label{tab:surf}}
\begin{tabular}{rrrrr}
$\log_{10}(r)$ & $\log_{10}(\Sigma_0)$& $\log_{10}(\epsilon_\Sigma)$ 
	& $\log_{10}(\Sigma_p)$& $\log_{10}(\epsilon_\Sigma)$ \\
\tableline
 -1.470&  0.613&  0.091&  0.426&  0.084\\
 -1.250&  0.529&  0.132&  0.497&  0.153\\
 -1.150&  0.410&  0.112&  0.440&  0.186\\
 -1.050&  0.387&  0.148&  0.340&  0.103\\
 -0.950&  0.182&  0.110&  0.282&  0.093\\
 -0.840&  0.188&  0.064&  0.018&  0.111\\
 -0.750& -0.011&  0.088& -0.101&  0.085\\
 -0.650&  0.049&  0.062&  0.020&  0.066\\
 -0.550& -0.144&  0.059& -0.118&  0.076\\
 -0.450& -0.188&  0.059& -0.219&  0.058\\
 -0.350& -0.436&  0.055& -0.442&  0.053\\
 -0.250& -0.673&  0.065& -0.559&  0.055\\
 -0.150& -0.679&  0.068& -0.664&  0.066\\
 -0.050& -0.836&  0.070& -0.846&  0.062\\
  0.050& -0.997&  0.074& -0.945&  0.084\\
  0.150& -1.293&  0.106& -1.187&  0.109\\
  0.250& -1.680&  0.309& -1.735&  0.372\\
  0.350& -2.053&  0.288& -3.632&  1.653\\
  0.440& -2.106&  0.448& -2.106&  0.448\\

\end{tabular}
\end{table}

\begin{table}
\caption{Line-of-Sight Velocity Dispersion (51 point)\label{tab:sig}}
\begin{tabular}{rrrr}
$R$ & $\sigma_p$ & $\sigma_p(-1\sigma)$ & $\sigma_p(+1\sigma)$ \\
\tableline
 0.10& 2.28& 1.95& 2.88\\
 0.20& 2.05& 1.68& 2.36\\
 0.30& 2.57& 1.69& 2.42\\
 0.40& 2.58& 1.72& 2.87\\
 0.50& 1.72& 1.38& 2.38\\
 0.60& 1.77& 1.05& 2.07\\
 0.70& 2.25& 1.38& 3.02\\
 0.80& 3.85& 2.52& 3.77\\
 0.90& 3.16& 1.46& 3.22\\
 1.00& 2.10& 1.56& 3.11\\
 1.10& 1.78& 1.27& 2.38\\
 1.20& 3.14& 1.62& 2.82\\
 1.30& 3.31& 1.17& 2.89\\
 1.40& 4.10& 0.96& 3.48\\
 1.51& 1.92& 0.00& 2.65\\
 1.61& 0.98& 0.00& 4.12\\

\end{tabular}
\end{table}

\begin{table}
\caption{Error Budget \label{tab:err}}
\begin{tabular}{llr}
Source of Error & Method & Error Estimate \\
\tableline
$M_v/L$ & jacknife & 17\% \\
$\rho_c/j$  & jacknife & 12\% \\
flattening correction  & jacknife & 4\% \\
$M_v$ bias ($\beta=0.5$)  & model fit & 17\% \\
$M_v$ bias ($0\le\beta\le 0.75$) & systematic & 10\% \\
$r_v$ normalization & bootstrap & 10\% \\
\sig1\ normalization & bootstrap & 8\% \\
$L_r^k$ bias & population average & 5\% \\
$L_r^k$ bias & systematic  & 10\% \\
\tableline
random errors & quadrature sum & 30\% \\
random errors & linear sum & 73\% \\
systematic errors & linear sum & 20\% \\
\end{tabular}
\end{table}

\clearpage

\figcaption[cvt.ps]{(NOT INCLUDED: see website)
The ensemble cluster and field constructed from the normalized sample,
deleting MS0906+11 and MS1358+62 which are ``binaries''. A total of
about 1500 galaxies are plotted.\label{fig:cvt}}

\figcaption[vdist.ps]{The background density as a function of
normalized velocity from the cluster center.  The horizontal line
shows the calculated mean value of the background over the interval 5
to 25\sig1. The dashed histogram is the lower histogram times 10. The
background for the full sample isin the left panel and for the sample
that excludes the low redshift clusters is in the right
panel.\label{fig:back}}

\figcaption[surf.ps]{The surface number density profile of
the ensemble cluster. The squares give the surface densities centered
about the peak density, the circles centered about the BCG center. The
line is the projection of the Hernquist function fit about the BCGs.
\label{fig:surf}}

\figcaption[sigfit.ps]{A fit to $\sigp(R)$ convolving
a radial velocity dispersion of the function
$\sigma_r^2(r)=(1+r)^{-1}$ with the fit to the surface density for
$\beta$ of 0, 0.25, 0.5, and 0.75, as lines of
increasing curvature. The error flags have been symmetrized and all
are independent. The jagged line is the 51 point smoothed velocity
dispersion. \label{fig:sigfit}}

\figcaption[sig.ps]{The 101 point moving average calculation of
the projected velocity dispersion. The model lines are for a
light-traces-mass model, with $\beta=\onehalf$ (solid) and $\beta=0,1$
(at large radii, the upper and lower dashed lines, respectively). The
line at the bottom is the absolute value of the local mean
velocity. \label{fig:sig}}

\figcaption[mlrad.ps]{The $M_v/L$ calculated from data of limited
sky coverage on a cluster. The lower line is for
background subtracted data. \label{fig:mlrad}}

\figcaption[mol.ps]{The ratio of the derived $M(r)$ to $L(r)$
for $\beta$ of 0, 0.25, 0.5, and 0.75 (strongly
circular to completely radial), from top to bottom at small radius,
respectively for $M_r^k=-18.5$ mag. The statistical errors are about
15\%. \label{fig:mol}}

\figcaption[mvb.ps]{The virial mass bias, $b_{Mv}$, at
\r200\ for varying $\beta$. The symbols denote
absolute magnitude cutoffs of -18.5 (triangles), -19.0 (squares) and
-19.5 (circles). The offset points with the dashed error flags are
about the peak of the number density.  \label{fig:mvb}}

\figcaption[mvg.ps]{The gradient of the virial
mass bias, $b_{Mv}$ as a function of $\beta$.  The symbols denote
absolute magnitude cutoffs of -18.5 (triangles), -19.0 (squares) and
-19.5 (circles) which indicates the small changes for slightly
different fits to the observed distributions. The points offset in
$\beta$ with dashed error flags are fitted about the peak of the
number density. 
\label{fig:mvg}}

\figcaption[czr.ps]{(NOT INCLUDED: see website)
The redshift corrected
color vs dimensionless redshift-radius. The BCGs are
plotted at a uniform radius of 0.09 units. The average
in bins of 0.2 radius unit is plotted as the solid line.\label{fig:czr}}

\figcaption[lzr.ps]{The k-corrected luminosity vs
redshift-radius. The average above the luminosity limit  is plotted as
a solid line.
\label{fig:lzr}}

\begin{figure}[h] \epsscale{0.6}\figurenum{2}\plotone{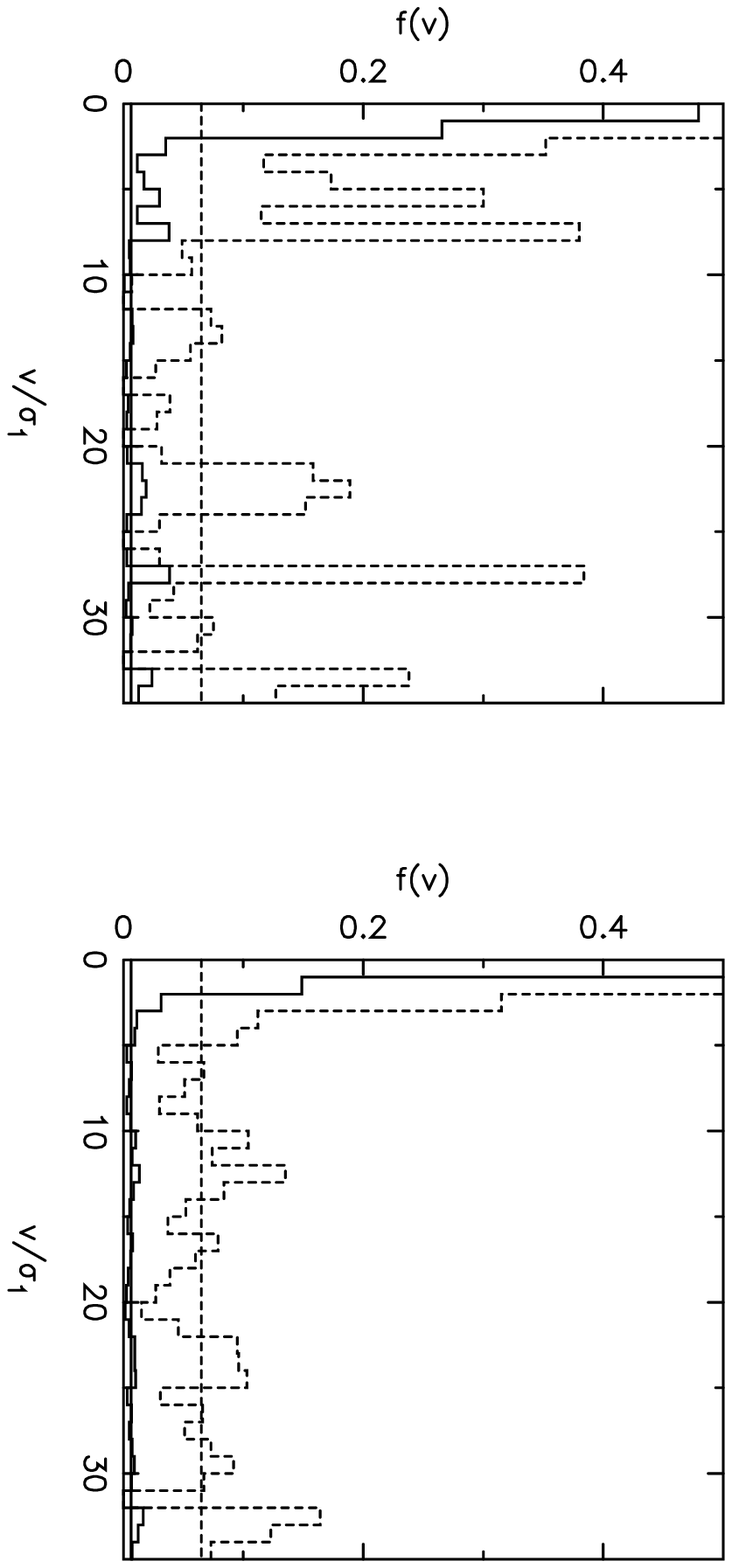} 
	\caption{}\end{figure}  \epsscale{1.0}
\begin{figure}[h] \figurenum{3}\plotone{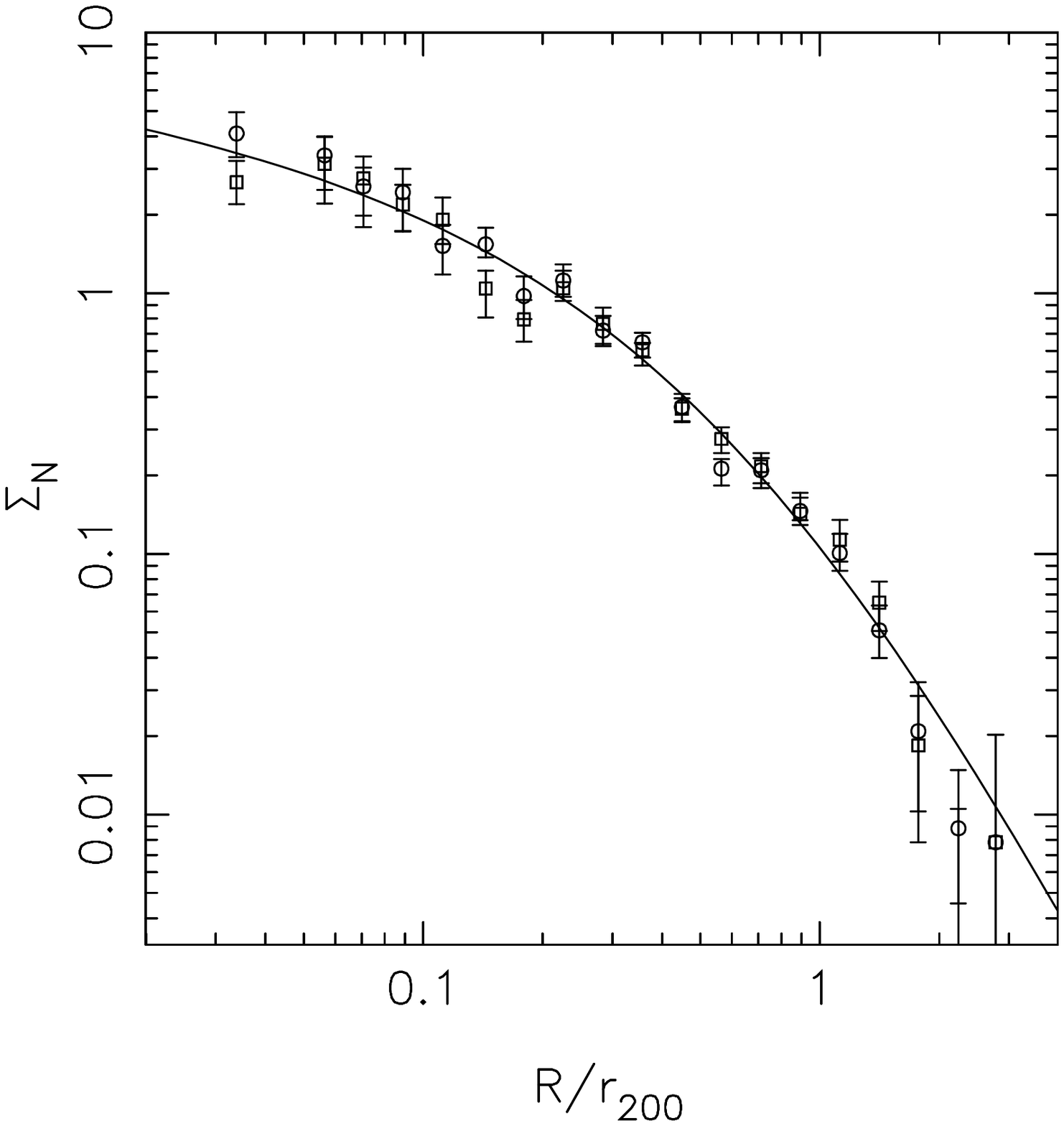} \caption{}\end{figure}  
\begin{figure}[h] \figurenum{4}\plotone{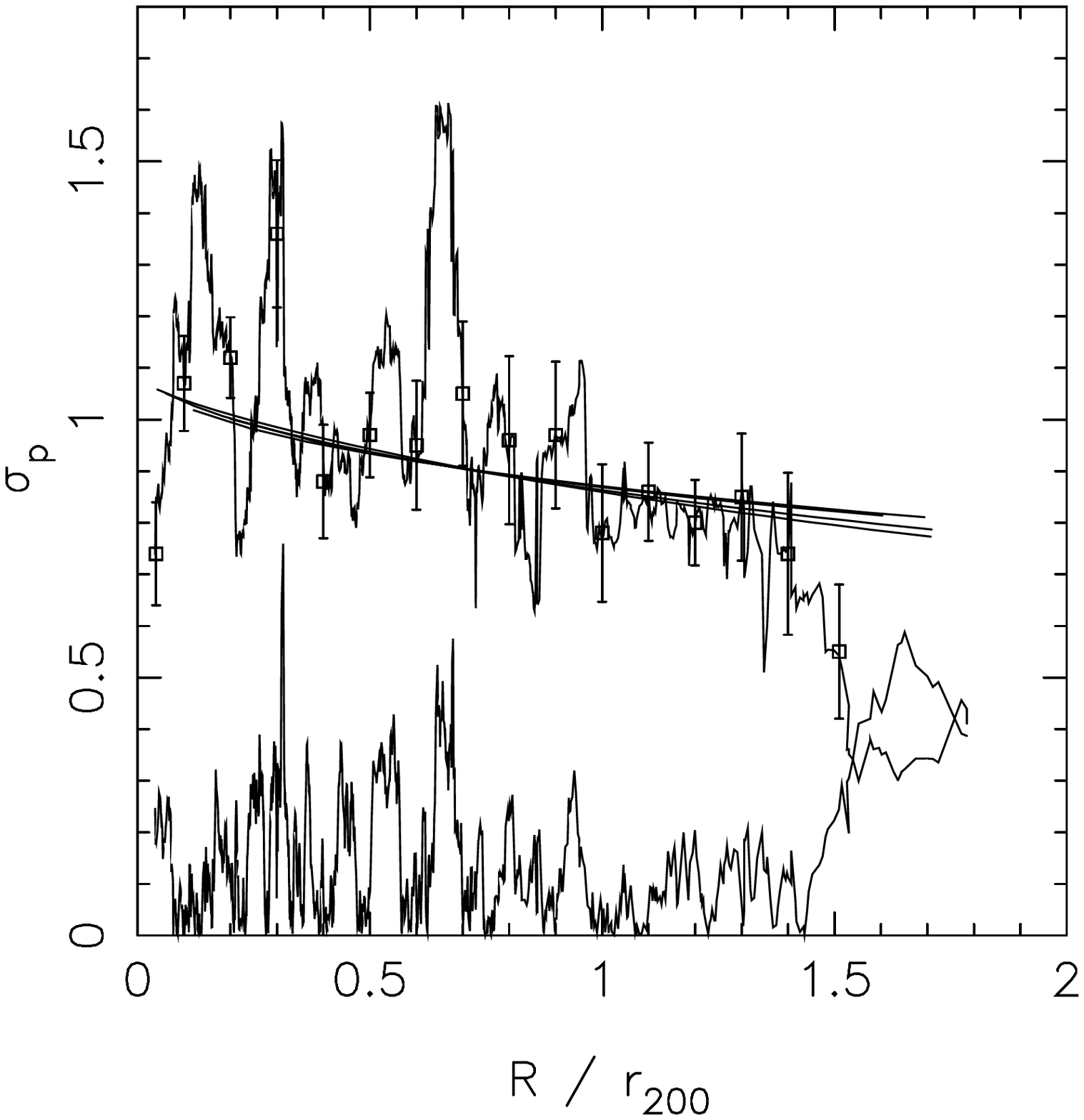} \caption{}\end{figure}  
\begin{figure}[h] \figurenum{5}\plotone{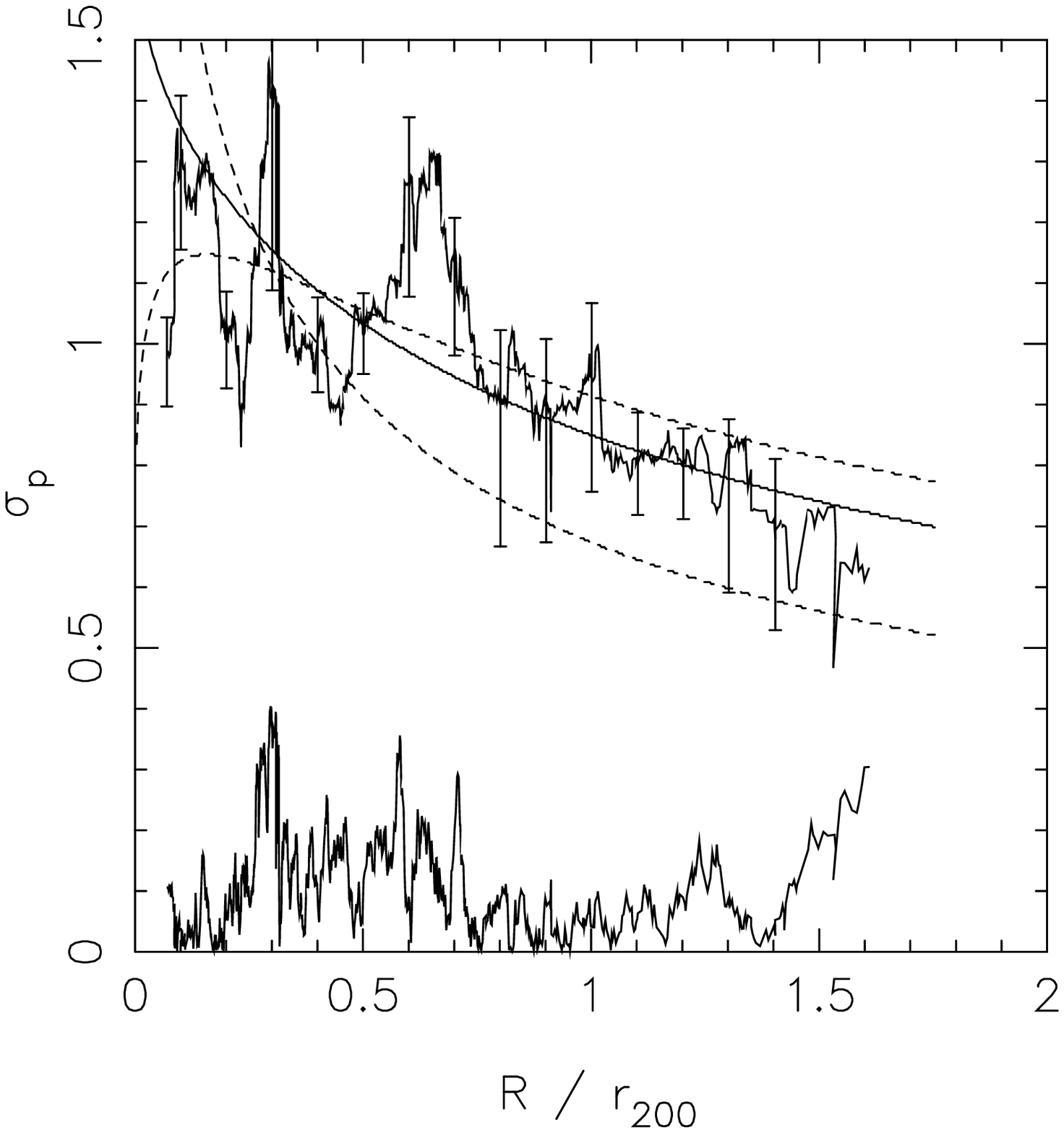} \caption{}\end{figure}  
\begin{figure}[h] \figurenum{6}\plotone{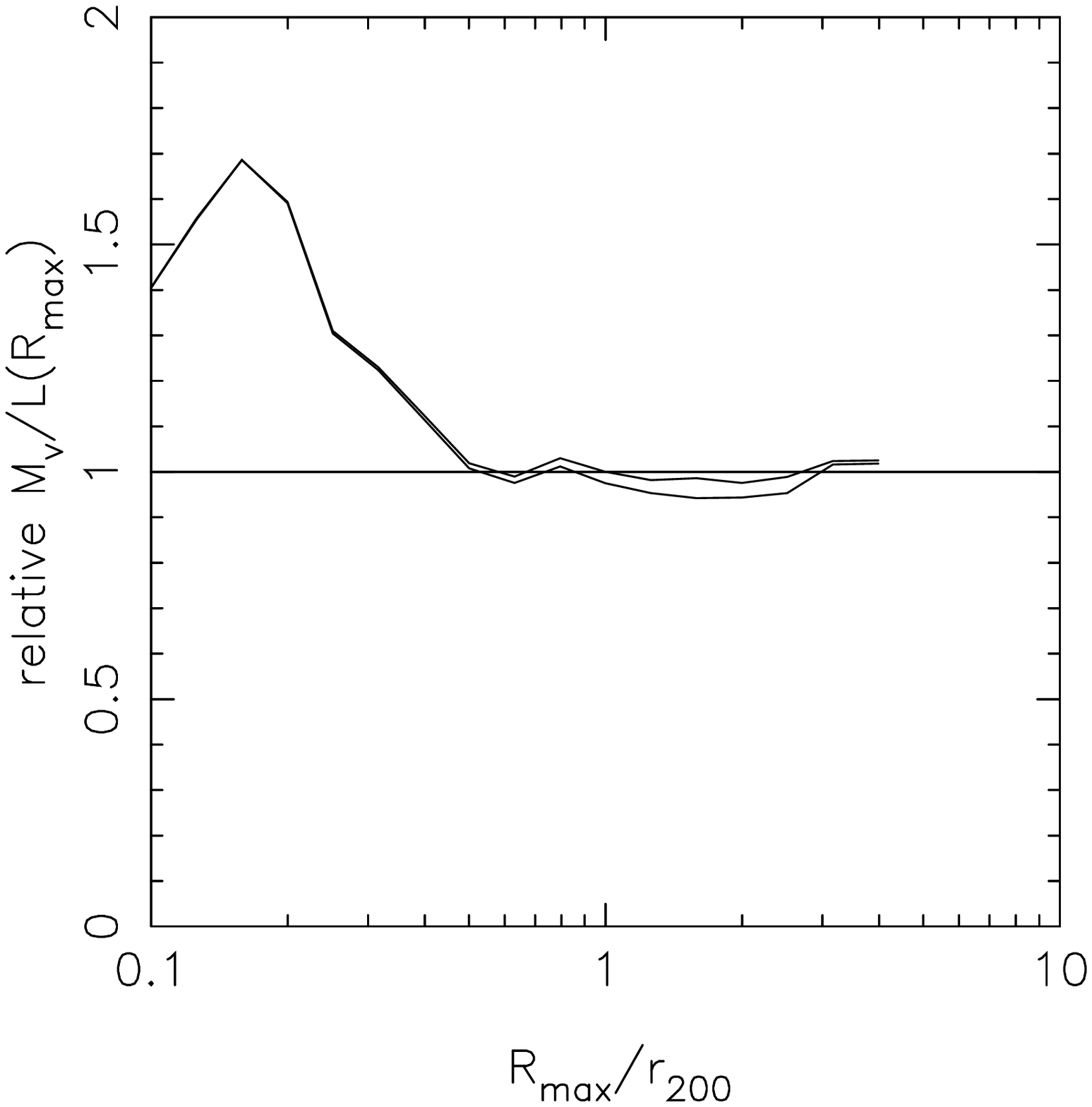} \caption{}\end{figure}  
\begin{figure}[h] \figurenum{7}\plotone{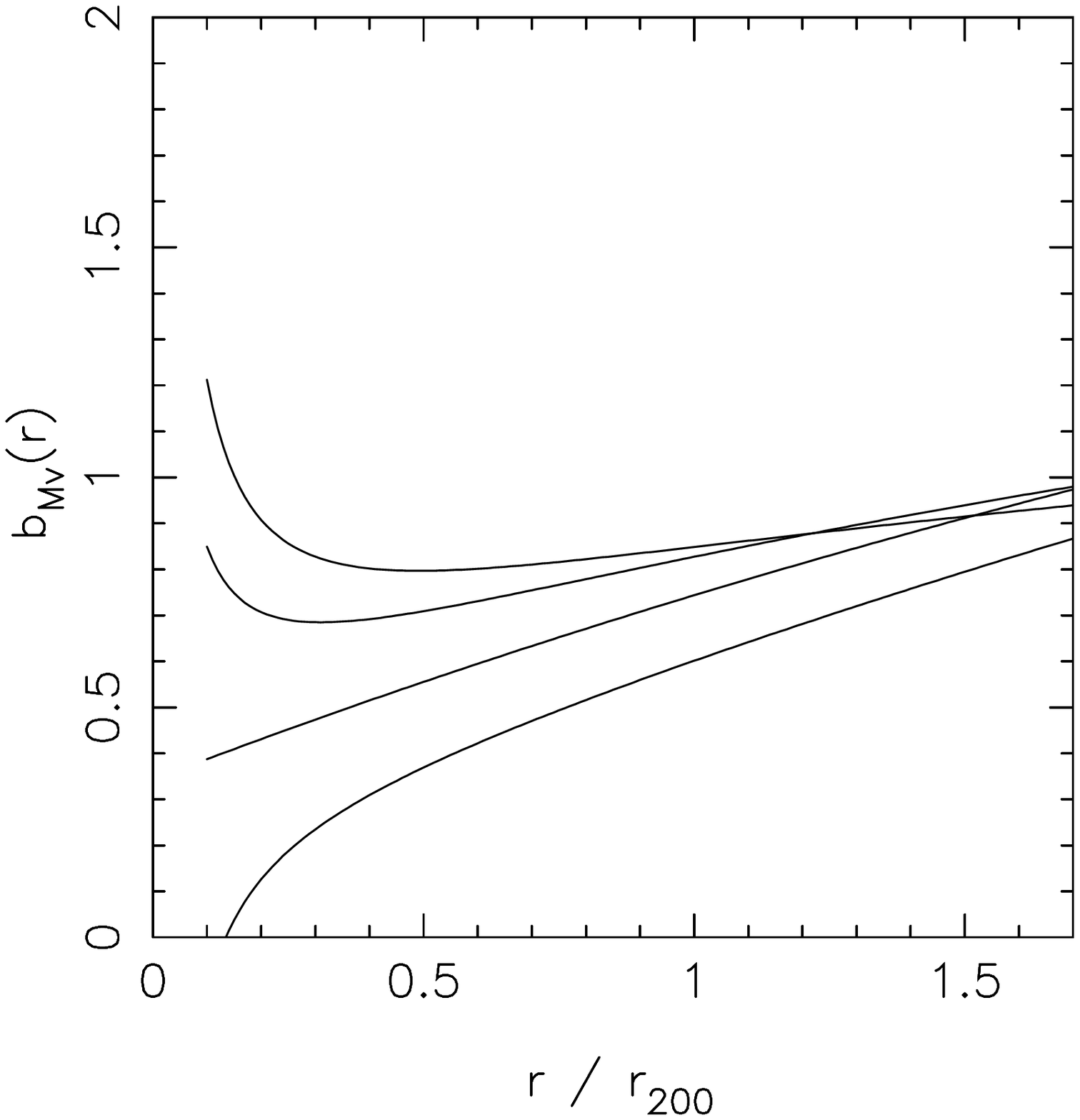} \caption{}\end{figure}  
\begin{figure}[h] \figurenum{8}\plotone{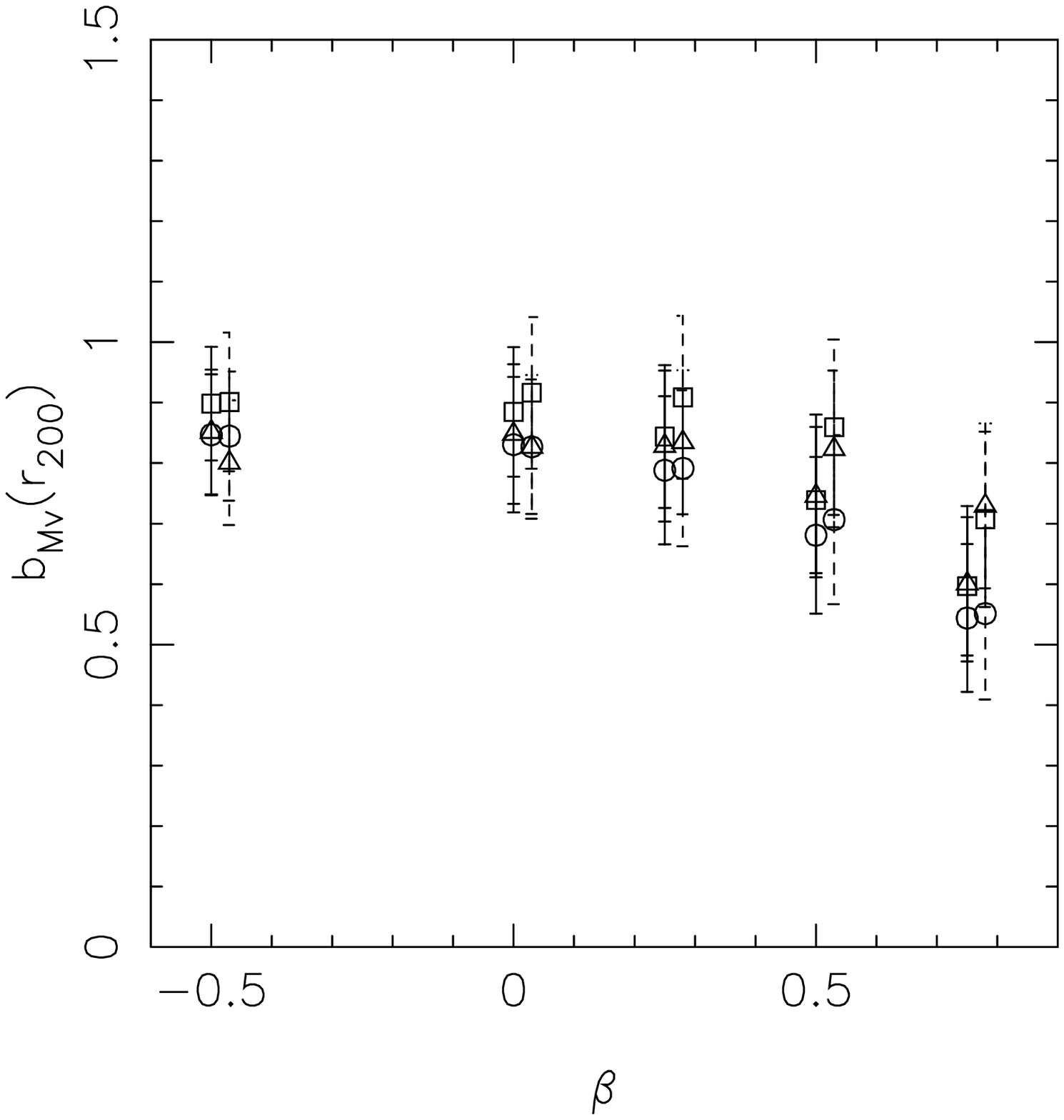} \caption{}\end{figure}  
\begin{figure}[h] \figurenum{9}\plotone{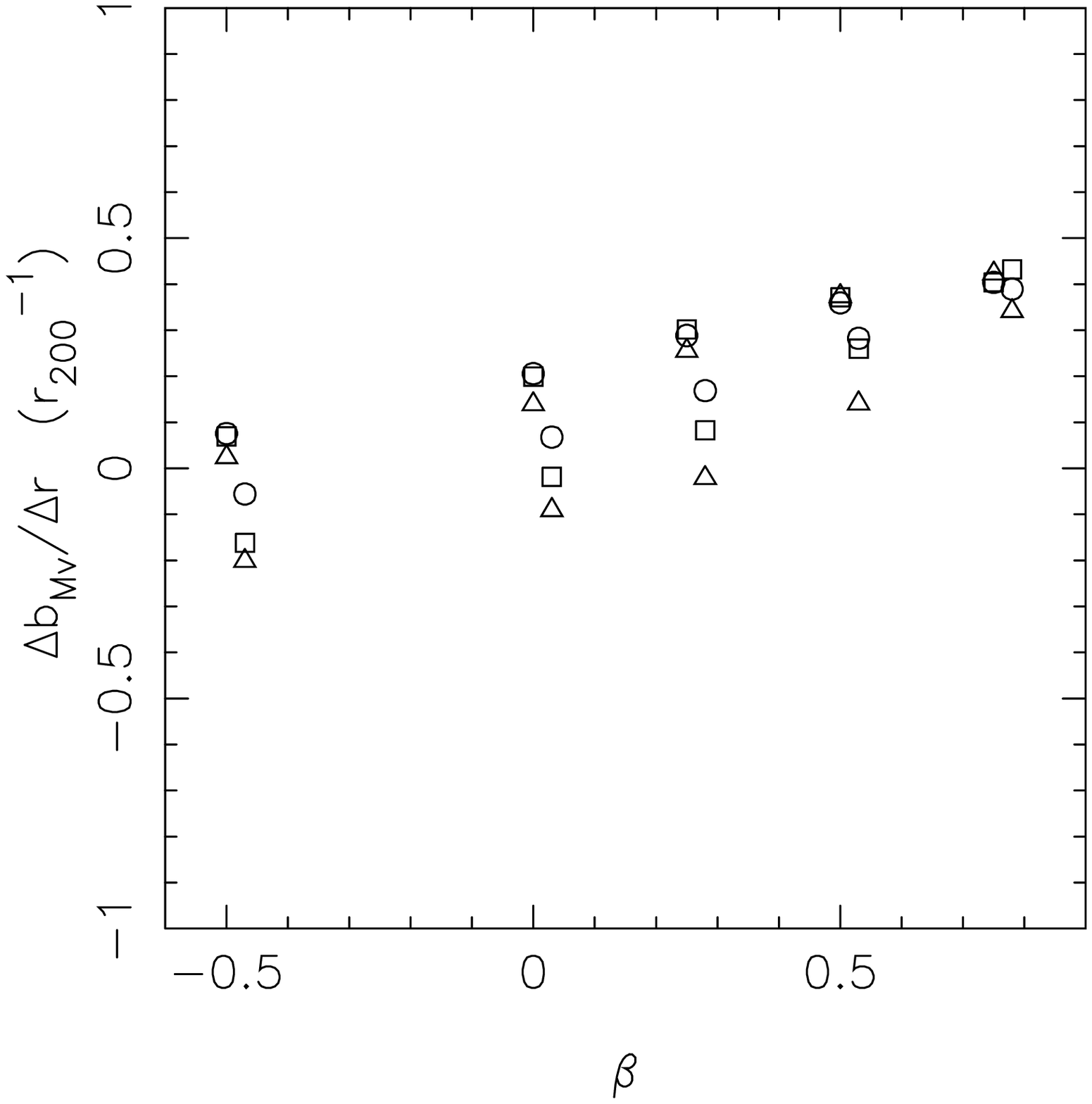} \caption{}\end{figure}  
\begin{figure}[h] \figurenum{11}\plotone{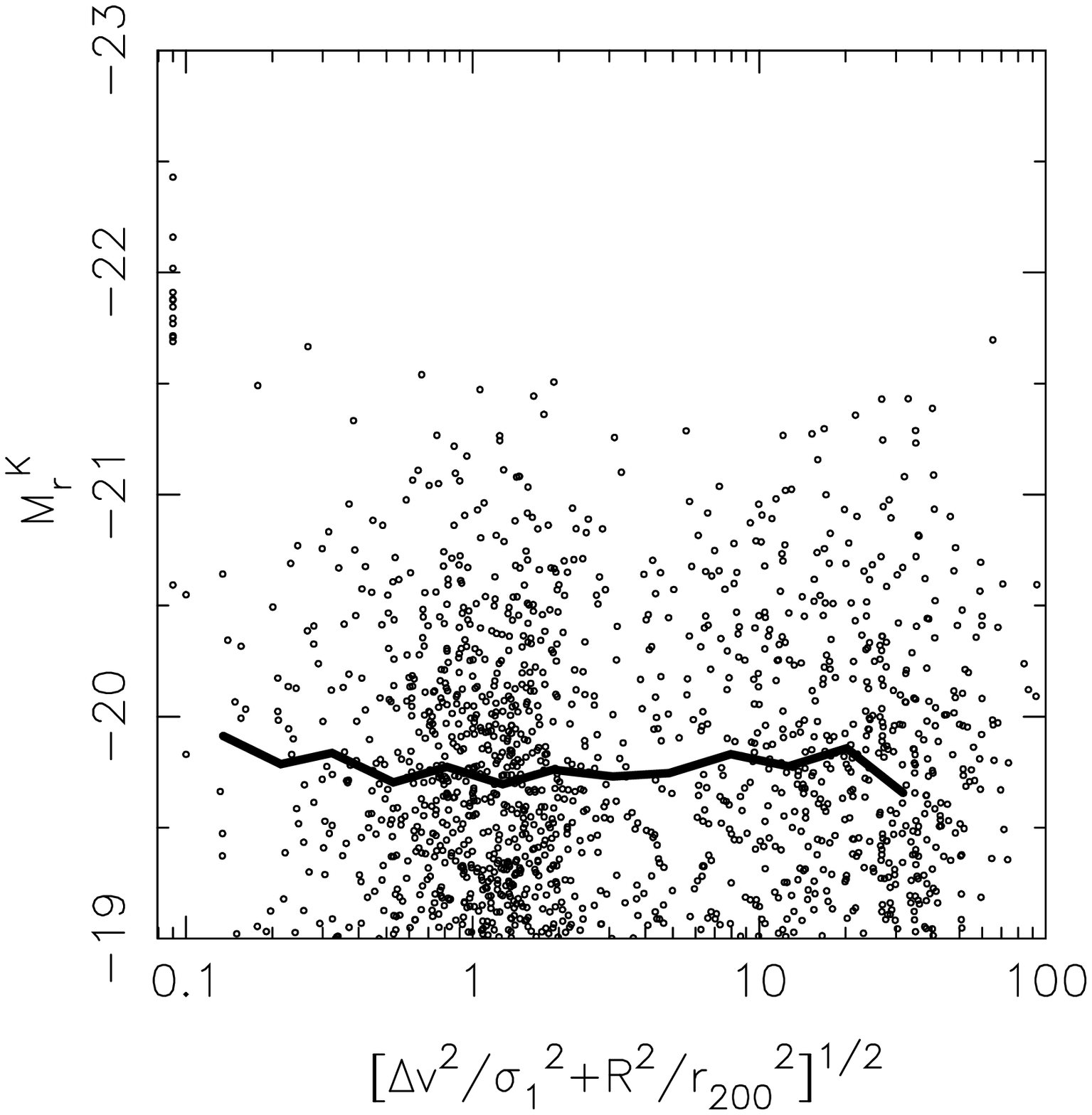} \caption{}\end{figure}  


\begin{thebibliography}{DUM}

\bibitem[Abell 1965]{abellrev}
Abell, G. O. 1965,  \araa, 3, 1
\bibitem[Abraham \et\ 1996]{a2390}
Abraham, R. G., Smecker-Hane, T. A., Hutchings, J. B.,
	Carlberg, R. G., Yee, H. K. C., Ellingson, E., Morris, S.,
	Oke, J. B., Davidge, T. 1996, \apjs, in press
\bibitem[Bahcall \& Tremaine 1981]{bt81}
Bahcall, J. \& Tremaine, S. D. 1981, \apj, 244, 805
\bibitem[Barnes, Goodman \& Hut 1986]{bgh86}
Barnes, J., Goodman, J., \& Hut, P. 1986, \apj, 300, 112
\bibitem[Bean \et\ 1983]{bean}
Bean, A. J., Efstathiou, G., Ellis, R. S., Peterson, B. A. \& Shanks, T. 1983,
	\mnras, 205, 605
\bibitem[Beers, Flynn \& Gebhardt 1990]{bfg}
Beers, T. C., Flynn, K. \& Gebhardt, K. 1990, \aj, 100, 32
\bibitem[Bertschinger 1985]{bert}
Bertschinger, E. 1985, \apjs, 58, 39
\bibitem[Binney \& Tremaine 1987]{bt_gd}
Binney, J. \& Tremaine, S. 1987, {\it Galactic Dynamics}, (Princeton 
        University Press)
\bibitem[Bird 1994]{bird_bcg}
Bird, C. M. 1994, \aj, 107, 1637
\bibitem[Butcher \& Oemler 1984]{bo}
Butcher, H. \& Oemler, A. 1984, \apj, 285, 423
\bibitem[Carlberg \& Dubinski 1991]{cd}
Carlberg, R. G. \& Dubinski, J. 1991, \apj, 369, 13
\bibitem[Carlberg 1994]{vbias}
Carlberg, R. G. 1994, \apj, 433, 468
\bibitem[Carlberg \et\ 1994]{cnoc1}
Carlberg, R. G., Yee, H. K. C., Ellingson, E.,
         Pritchet C., Abraham, R., Smecker-Hane, T., Bond, J. R.,
        Couchman, H. M. P., Crabtree, D., 
        Crampton, D., Davidge, T., Durand, D.,
        Eales, S., Hartwick, F. D. A., Hesser, J. E.,
        Hutchings, J. B., Kaiser, N., Mendes de Oliveira, C.,
        Myers, S. T., Oke, J. B., Rigler, M. A., Schade, D.,
        \& West, M. 1994, \jrasc, 88, 39
\bibitem[C96]{global}
Carlberg, R. G., Yee, H. K. C., Ellingson, E., Abraham, R.,
	Gravel, P., Morris, S. M, \& Pritchet, C. J. 1996a (C96), 
	\apj, 462, 32
\bibitem[Cole \& Lacey 1996]{cole_lacey}
Cole, S. \& Lacey, C. 1996, \mnras, 281, 716
\bibitem[Copi, Schramm \& Turner 1995]{copi}
Copi, C. J., Schramm, D. N., \& Turner, M. S. 1995, Science, 267, 192
\bibitem[Crone, Evrard \& Richstone 1994]{cer}
Crone, M. M., Evrard, A. E., \& Richstone, D. O. 1994, \apj, 434, 402
\bibitem[Davis \& Peebles 1983]{dp}
Davis, M. \& Peebles, P. J. E. 1983, \apj, 267, 465
\bibitem[Dekel 1994]{dekel}
Dekel, A. 1994, \araa, 32, 371
\bibitem[Dekel \et\ 1993]{potiras}
Dekel, A., Bertschinger, E., Yahil, A., Strauss, M. A.,
        Davis, M., \& Huchra, J. P. 1993, \apj, 412, 1
\bibitem[Dubinski 1993]{dubinski}
Dubinski, J. 1993, \apj, 401, 441
\bibitem[Efron \& Tibshirani 1986]{et}
Efron, B. \& Tibshirani, R. 1986, {\it Statistical Science}, 1, 54
\bibitem[Efstathiou, Ellis \& Peterson 1988]{eep} 
Efstathiou, G., Ellis, R. S., \& Peterson, B. A. 1988, 
        \mnras, 232, 431
\bibitem[Fillmore \& Goldreich 1984]{fg}
Fillmore, J. A. \& Goldreich, P. 1984, \apj, 281, 1
\bibitem[Fisher \et\ 1994]{iras}
Fisher, K. B., Davis, M., Strauss, M. A., Yahil, A., \& 
        Huchra, J. P. 1994, \mnras, 267, 927
\bibitem[Gioia \& Luppino 1994]{gl}
Gioia, I. M. \& Luppino, G. A. 1994, \apjs, 94, 583
\bibitem[Gioia \et\ 1990]{emss1}
Gioia, I. M., Maccacaro, T., Schild, R. E., Wolter,
        Stocke, J. T., Morris, S. L., \& Henry, J. P. 1990, \apjs, 72, 567
\bibitem[Gott \& Gunn 1972]{gg}
Gott, J. R. \& Gunn, J. 1972, \apj, 176, 1
\bibitem[Gott \& Turner 1976]{gt}
Gott, J. R. \& Turner, E. L 1976, \apj, 209, 1
\bibitem[Gunn 1978]{gunn}
Gunn, J. E. 1978, in {\it Observational Cosmology}, eds. Maeder, A.,
        Martinet, L, \& Tammann, G. 1978 (Geneva Observatory: 
        Sauverny)
\bibitem[den Hartog 1995]{hartog}
den Hartog, R. 1995, {\it The Dynamics of Rich Galaxy Clusters},
	Ph.~D. Thesis, Leiden University
\bibitem[Henry \et\ 1992]{emss2}
Henry, J. P., Gioia, I. M., Maccacaro, T., Morris, S. L.,
	Stocke, J. T., \& Wolter, A. 1992, \apj, 386, 408
\bibitem[Hernquist 1990]{hernquist}
Hernquist, L. 1990, \apj, 356, 359
\bibitem[Kaiser 1987]{kaiser}
Kaiser, N. 1987, \mnras, 227, 1
\bibitem[van Kampen 1995]{vank}
van Kampen, E. 1995, \mnras, 273, 295
\bibitem[Katz \& White 1993]{kw}
Katz, N. \& White, S. D. M. 1993, \apj, 412, 455
\bibitem[Kent \& Gunn 1982]{kg}
Kent, S. \& Gunn, J. E. 1982, \aj, 87, 945
\bibitem[LeF\`evre \et\ 1994]{mos}
LeF\`{e}vre, O., Crampton, D., Felenbok, P., \& Monnet, G. 1994, \aap, 282, 340
\bibitem[Limber 1959]{limber}
Limber, D. N. 1959, \apj, 130, 414
\bibitem[Loveday \et\ 1993]{loveday} 
Loveday, J., Efstathiou, G., Peterson, B. A., 
        Maddox, S. J. 1993, \apj, 390, 338
\bibitem[Merritt 1984]{merritt}
Merritt, D. 1984, \apj, 313, 121
\bibitem[Merritt \& Tremblay 1994]{mt}
Merritt, D. \& Tremblay, B. 1994, \aj, 108, 514
\bibitem[Mohr \et\ 1995]{mohr}
Mohr, J. J., Evrard, A. E., Frabicant, D. G., \& Geller, M. J. 1995,
	\apj, 447, 8
\bibitem[Navarro, Frenk \& White 1995]{nfw}
Navarro, J. F., Frenk, C. S., \& White, S. D. M. 1995, ApJ submitted
\bibitem[Oort 1958]{oort}
Oort, J. H. 1958, in {\it La Structure et L'\'Evolution de L'Univers},
	Onzi\`eme Conseil de Physique, ed. R. Stoops (Solvay
	Institute: Bruxelles) p. 163
\bibitem[Palmer \& Papaloizou 1987]{pp87}
Palmer, P. L. \& Papaloizou, J. 1987, \mnras, 224, 1043
\bibitem[Peebles 1970]{peebles_coma}
Peebles, P. J. E. 1970, \aj, 75, 13
\bibitem[Peebles 1993]{ppc}
Peebles, P. J. E. 1993, {\it Principles of Physical Cosmology}
	(Princeton University Press: Princeton)
\bibitem[Press \et\ 1992]{nr}
Press, W. H., Teukolsky, S. A., Vetterling, W. T., \& Flannery, B. P. 1992,
	{\it Numerical Recipes in C} (Cambridge University Press)
\bibitem[Pryor \& Geller 1984]{pg84}
Pryor, C. \& Geller, M. J. 1984, \apj, 278, 457
\bibitem[Ramella, Geller \& Huchra 1989]{rgh}
Ramella, M., Geller, M. J., \& Huchra, J. P. 1989, \apj, 344, 57
\bibitem[Reg\"os \& Geller 1989]{regos_geller}
Reg\"os, A. \& Geller, M. J. 1989, \aj, 98, 755
\bibitem[Richstone, Loeb \& Turner 1992]{rlt}
Richstone, D. O., Loeb, A., \& Turner, E. L. 1992, \apj, 393, 477
\bibitem[Richstone \& Tremaine 1984]{rt84}
Richstone, D. O. \& Tremaine, S. T. 1984, \apj, 286, 27
\bibitem[Rood \et\ 1972]{rpkk}
Rood, H. J., Page, T. L, Kintner, E. C. \& King, I. R. 1972,
        \apj, 175, 627
\bibitem[Schade \et\ 1996a]{schade_e}
Schade, D., Carlberg, R. G., Yee,
        H. K. C., L\'opez-Cruz, O. \& Ellingson, E. 1996a, \apjl, 464, L63
\bibitem[1996b]{schade_d}
Schade, D., Carlberg, R. G., Yee,
        H. K. C., L\'opez-Cruz, O. \& Ellingson, E. 1996b, \apjl, 465, L103
\bibitem[Schwarzschild 1954]{schwarz}
Schwarzschild, M. 1954, \aj, 59, 273
\bibitem[Shaya, Peebles \& Tully 1995]{spt}
Shaya, E. J., Peebles, P. J. E., \& Tully, R. B. 1995, \apj, 454, 15
\bibitem[Smith 1936]{smith}
Smith, S. 1936, \apj, 83, 29
\bibitem[Strauss \& Willick 1995]{sw}
Strauss, M. A. \& Willick, J. A. 1995, \physrep, 261, 271
\bibitem[The \& White 1986]{tw84}
The, L. S. \& White, S. D. M. 1984, \aj, 92, 1248
\bibitem[Tremaine \et\ 1994]{eta}
Tremaine, S., Richstone, D. O., Byun, Y-I., Dressler, A.,
        Faber, S. M., Grillmair, C., Kormendy, J., \& Lauer, T. R. 1994,
        \aj, 107, 634
\bibitem[Tsai \& Buote 1995]{tsai_buote}
Tsai, J. C. \& Buote, D. A. 1996, \apj, 458, 27
\bibitem[van Albada 1960]{vanA}
van Albada, G. B. 1960, \bain, 15, 165
\bibitem[Walker \et\ 1991]{walker}
Walker, T. P., Steigman, G., Kang, H., Schramm, D. M. \& Olive, K. A.
    1991, \apj, 376, 51
\bibitem[West \& Richstone 1988]{wr}
West, M. J. \& Richstone, D. O. 1988, \apj, 335, 532
\bibitem[White \& Fabian 1995]{wf}
White, D. A. \& Fabian, A. C. 1995, \mnras, 273, 72
\bibitem[White 1976]{white_coma}
White, S. D. M. 1976, \mnras, 177, 717
\bibitem[White \et\ 1993]{wnef}
White, S. D. M.,  Navarro, J. F., Evrard, A. E., \& 
	Frenk, C. S. 1993, \nat, 366, 429
\bibitem[Yee, Ellingson \& Carlberg 1996]{yec}
Yee, H. K. C., Ellingson, E. \& Carlberg, R. G. 1996, \apjs, 102, 269
\bibitem[Yee \et\ 1995]{cnoc_bo}
Yee, H. K. C., Sawicki, M. J., Ellingson, E., \& Carlberg, R. G. 1995,
	in {\it Fresh Views on Elliptical Galaxies} ASP conference series,
	86, 301
\bibitem[Zembrowski \& Carlberg 1996]{zembrowski}
Zembrowski, P. \& Carlberg, R. G. 1996, in preparation
\bibitem[Zwicky 1933]{z33}
Zwicky, F. 1933, Helv. Phys. Acta 6, 110
\bibitem[Zwicky 1937]{z37}
Zwicky, F. 1937, \apj, 86, 217

\end{thebibliography}
\end{document}